\documentclass[trackchanges,twocolumn, twocolappendix]{aastex701}

\newcommand{\gsim}{\raisebox{-0.13cm}{~\shortstack{$>$ \\[-0.07cm]
      $\sim$}}~}

\usepackage{amsmath}
\usepackage{tikz-feynman}
\DeclareMathOperator*{\argmin}{argmin}

\usetikzlibrary{shapes.geometric, arrows, calc}
\tikzstyle{startstop} = [rectangle, rounded corners, minimum width=1cm, minimum height=1cm,text centered, draw=white]
\tikzstyle{arrow} = [thick,->,>=stealth,]

\begin{document}

\title{Strong Gravitational Lensing Posterior Sampling in Pixel-Space Using Diffusion Models and Recurrent Inference Machines}

\author[orcid=0000-0002-0997-4827]{Guillaume Payeur}
\affiliation{Department of Physics, Universit\'e de Montr\'eal, Montreal QC, Canada}
\affiliation{Mila - Quebec AI Institute, Montreal QC, Canada}
\affiliation{Ciela Institute, Institute for Astrophysics and Machine Learning, Montreal QC, Canada}
\affiliation{Trottier Space Institute, McGill University, Montreal QC, Canada}
\email[show]{guillaume.payeur@umontreal.ca}

\author[orcid=0000-0003-3544-3939]{Laurence Perreault-Levasseur}
\affiliation{Department of Physics, Universit\'e de Montr\'eal, Montreal QC, Canada}
\affiliation{Mila - Quebec AI Institute, Montreal QC, Canada}
\affiliation{Ciela Institute, Institute for Astrophysics and Machine Learning, Montreal QC, Canada}
\affiliation{Trottier Space Institute, McGill University, Montreal QC, Canada}
\affiliation{Center for Computational Astrophysics, Flatiron Institute, New York, USA}
\affiliation{Perimeter Institute for Theoretical Physics, Waterloo ON, Canada}
\email{laurence.perreault.levasseur@umontreal.ca}

\author[orcid=0009-0008-5839-5937]{Gabriel Missael Barco}
\affiliation{Department of Physics, Universit\'e de Montr\'eal, Montreal QC, Canada}
\affiliation{Mila - Quebec AI Institute, Montreal QC, Canada}
\affiliation{Ciela Institute, Institute for Astrophysics and Machine Learning, Montreal QC, Canada}
\email{gabriel.missael.barco@umontreal.ca}

\author[orcid=0000-0002-8669-5733]{Yashar Hezaveh}
\affiliation{Department of Physics, Universit\'e de Montr\'eal, Montreal QC, Canada}
\affiliation{Mila - Quebec AI Institute, Montreal QC, Canada}
\affiliation{Ciela Institute, Institute for Astrophysics and Machine Learning, Montreal QC, Canada}\
\affiliation{Trottier Space Institute, McGill University, Montreal QC, Canada}
\affiliation{Center for Computational Astrophysics, Flatiron Institute, New York, USA}
\affiliation{Perimeter Institute for Theoretical Physics, Waterloo ON, Canada}
\email{yashar.hezaveh@umontreal.ca}

\begin{abstract}

Modeling galaxy-galaxy strong gravitational lenses to infer the brightness of the source galaxy and the mass distribution of the foreground galaxy is computationally challenging, particularly for high-resolution, high signal-to-noise ratio observations. In this regime, high-dimensional representations of both the source and the foreground mass distribution are necessary to model the data down to the noise level. This inference problem has been challenging for both traditional and machine learning-based methods because of its high dimensionality and its non-linearity in the foreground mass distribution. We present a method to generate joint posterior samples of the source galaxy and foreground mass distribution as pixelated images conditioned on observations. The method combines diffusion-based generative modeling and recurrent inference machines. It can model realistic gravitational lensing simulations with background and foreground galaxies drawn from cosmological hydrodynamical simulations down to the noise level. 

\end{abstract}



\section{Introduction}

Galaxy-galaxy strong gravitational lensing is the physical process wherein the gravitational potential of a foreground galaxy bends the trajectory of light emitted by a distant background galaxy in a way that multiple distorted images of the source are observed from Earth. Observations of these systems can answer many questions in cosmology and astrophysics. As an example, the distortions of the images of the source are tracers of the mass distribution of the foreground galaxy, allowing for the constraint of the properties of dark matter on galactic scales (see, e.g., \citealt{
Mao_Schneider_1998,
Dalal_Kochanek_2002, 
Treu_Koopmans_2004, 
Richard_Smith_Kneib_Ellis_Sanderson_Pei_Targett_Sand_Swinbank_Dannerbauer_etal._2010, Vegetti_Koopmans_Bolton_Treu_Gavazzi_2010, Vegetti_Koopmans_Auger_Treu_Bolton_2014,
Grillo_Suyu_Rosati_Mercurio_Balestra_Munari_Nonino_Caminha_Lombardi_DeLucia_etal._2015,
Hezaveh_Dalal_Marrone_Mao_Morningstar_Wen_Blandford_Carlstrom_Fassnacht_Holder_etal._2016,
Hsueh_Enzi_Vegetti_Auger_Fassnacht_Despali_Koopmans_McKean_2020,
Gilman_Birrer_Nierenberg_Treu_Du_Benson_2020, 
Gilman_Bovy_Treu_Nierenberg_Birrer_Benson_Sameie_2021,
Minor_Gad-Nasr_Kaplinghat_Vegetti_2021,
AnauMontel_Coogan_Correa_Karchev_Weniger_2022, 
Bayer_Chatterjee_Koopmans_Vegetti_McKean_Treu_Fassnacht_Glazebrook_2023, 
Bayer_Koopmans_McKean_Vegetti_Treu_Fassnacht_Glazebrook_2023, 
Powell_Vegetti_McKean_White_Ferreira_May_Spingola_2023,
Ballard_Enzi_Collett_Turner_Smith_2024} and see \citealt{Vegetti_Birrer_Despali_Fassnacht_Gilman_Hezaveh_PerreaultLevasseur_McKean_Powell_ORiordan_etal._2024} for a review article). As another example, by considering the time delay between the multiple images, it is possible to constrain the rate of expansion of the universe given by the Hubble parameter (see, e.g., \citealt{Shajib_Birrer_Treu_Agnello_Buckley-Geer_Chan_Christensen_Lemon_Lin_Millon_etal._2020, Qi_Zhao_Cao_Biesiada_Liu_2021, Shajib_Mozumdar_Chen_Treu_Cappellari_Knabel_Suyu_Bennert_Frieman_Sluse_etal._2023, Colaco_Holanda_Santana_Silva_2025} and see \citealt{Birrer_Millon_Sluse_Shajib_Courbin_Erickson_Koopmans_Suyu_Treu_2024} for a review article). As a last example, one can study galaxy formation at high redshifts by observing magnified images of very distant galaxies which would otherwise be unresolved or too faint to be observed (see, e.g., \citealt{
Zitrin_Moustakas_Bradley_Coe_Moustakas_Postman_Shu_Zheng_Benitez_Bouwens_etal._2012,
Coe_Zitrin_Carrasco_Shu_Zheng_Postman_Bradley_Koekemoer_Bouwens_Broadhurst_etal._2013,
Stark_Walth_Charlot_Clement_Feltre_Gutkin_Richard_Mainali_Robertson_Siana_etal._2015,
Marques-Chaves_Perez-Fournon_Shu_Martinez-Navajas_Bolton_Kochanek_Oguri_Zheng_Mao_Montero-Dorta_etal._2017,
Marques-Chaves_Perez-Fournon_Gavazzi_Martinez-Navajas_Riechers_Rigopoulou_Cabrera-Lavers_Clements_Cooray_Farrah_etal._2018,
Fan_Wang_Yang_Keeton_Yue_Zabludoff_Bian_Bonaglia_Georgiev_Hennawi_etal._2019, 
Jacobs_Collett_Glazebrook_McCarthy_Qin_Abbott_Abdalla_Annis_Avila_Bechtol_etal._2019, 
Marques-Chaves_Perez-Fournon_Shu_Colina_Bolton_alvarez-Marquez_Brownstein_Cornachione_Geier_Jimenez-angel_etal._2020, 
Shu_Yang_Liu_Wang_Wang_Han_Huang_Lim_Chang_Zheng_etal._2022,
Shu_Canameras_Schuldt_Suyu_Taubenberger_Inoue_Jaelani_2022,
Nightingale_Mahler_McCleary_He_Hogg_Amvrosiadis_Gozaliasl_Mercier_Scognamiglio_Berman_etal._2025} and see \citealt{Shajib_Vernardos_Collett_Motta_Sluse_Williams_Saha_Birrer_Spiniello_Treu_2024} for a review article).

The scientific potential of galaxy-galaxy strong gravitational lensing can be realized via lens modeling, which is the process of jointly inferring the brightness distribution of the background source and the mass distribution of the foreground galaxy using observations. Traditionally, due to the computational cost and complexity of the inference problem, this has been done assuming that the source brightness and foreground mass distribution follow simple low-dimensional parametric profiles (e.g., \citealt{
Bolton_Burles_Koopmans_Treu_Gavazzi_Moustakas_Wayth_Schlegel_2008, 
Hezaveh_Levasseur_Marshall_2017, 
Savary_Rojas_Maus_Clement_Courbin_Gavazzi_Chan_Lemon_Vernardos_Canameras_etal._2022,
Knabel_Holwerda_Nightingale_Treu_Bilicki_Brough_Driver_Finnerty_Haberzettl_Hegde_etal._2023,
Huang_Baltasar_Ratier-Werbin_Storfer_Sheu_Agarwal_Tamargo-Arizmendi_Schlegel_Aguilar_Ahlen_etal._2026}). These simplistic assumptions often bias the inference   
and prevent the desired information from being extracted from the data \citep{Xu_Sluse_Gao_Wang_Frenk_Mao_Schneider_Springel_2015, Sonnenfeld_2018, Galan_Vernardos_Minor_Sluse_VanDeVyvere_Gomer_2024}. This is particularly true for high-resolution and high signal-to-noise ratio (SNR) observations.

In recent years, advances in machine learning have given rise to models that offer an alternative to traditional inference methods. On the one hand, diffusion-based generative models are a class of machine learning models that can generate samples from complex probability distributions $p(\boldsymbol{x})$, including in settings where $\boldsymbol{x}$ is high-dimensional and $p(\boldsymbol{x})$ is multi-modal \citep{Song_Ermon_2019,  Ho_Jain_Abbeel_2020, Song_Sohl-Dickstein_Kingma_Kumar_Ermon_Poole_2020, Wong_Suyu_Chen_Rusu_Millon_Sluse_Bonvin_Fassnacht_Taubenberger_Auger_etal._2020, Lai_Song_Kim_Mitsufuji_Ermon_2025}. They have been used for Bayesian analysis of high-dimensional models in the sciences, including in physics  (e.g., \citealt{ 
Song_Shen_Xing_Ermon_2021,
Adam_Coogan_Malkin_Legin_Perreault-Levasseur_Hezaveh_Bengio_2022,
Kazerouni_Aghdam_Heidari_Azad_Fayyaz_Hacihaliloglu_Merhof_2022,
Dia_Yantovski-Barth_Adam_Bowles_Lemos_Scaife_Hezaveh_Perreault-Levasseur_2023,
Legin_Ho_Lemos_Perreault-Levasseur_Ho_Hezaveh_Wandelt_2023,
Remy_Lanusse_Jeffrey_Liu_Starck_Osato_Schrabback_2023,
Xue_Li_Patel_Regier_2023,
Bourdin_Legin_Ho_Adam_Hezaveh_Perreault-Levasseur_2024,
Cuesta-Lazaro_Mishra-Sharma_2024,
Wu_Sun_Chen_Zhang_Yue_Bouman_2024,
Adam_Stone_Bottrell_Legin_Hezaveh_Perreaul-Levasseur_2025, 
Andry_Lewin_Rozet_Rochman_Mangeleer_Pirlet_Faulx_Gregoire_Louppe_2025, 
Barco_Adam_Stone_Hezaveh_Perreault-Levasseur_2025, 
Barco_Legin_Stone_Hezaveh_Perreault-Levasseur_2025, 
Boruah_Jacob_Jain_2025, 
Legin_Stone_Adam_Barco_Coogan_Malkin_Perreault-Levasseur_Hezaveh_2025, 
Riveros_Saavedra_Hortua_Garcia-Farieta_Olier_2025, 
Rozet_Andry_Lanusse_Louppe_2025, 
Stone_Legin_Adam_Malkin_Barco_Perreaul-Levasseur_Hezaveh_2025}). On the other hand, recurrent inference machines (RIMs) are a machine learning framework for solving inverse problems which proceeds by learning a data-driven iterative inference algorithm that implicitly captures the prior distribution over the model parameters. This framework has been applied to under-constrained inverse problems for which the prior distribution is hard to compute or intractable (e.g., \citealt{morningstar_analyzing_2018, morningstar_data-driven_2019, lonning_recurrent_2019, sabidussi_recurrent_2021, rhea_deconvolving_2024}), including to non-linear inverse problems (e.g., \citealt{Modi_Lanusse_Seljak_Spergel_Perreault-Levasseur_2021, Adam_Perreault-Levasseur_Hezaveh_Welling_2023}). In these works, the RIM produces a maximum a posteriori (MAP) estimate of the model parameters, given an observation.

Over the coming years, the Rubin Observatory Legacy Survey of Space and Time (LSST) and the Euclid Space Telescope are expected to discover over 100,000 strong gravitational lenses \citep{Collett_2015, Lines_Li_Collett_Holloway_Nightingale_Rojas_Verma_Walmsley_2025, Strong_Lensing_Science_Collaboration_2025}. Realizing the scientific potential of this data requires lens modeling methods that leverage flexible, high-dimensional representations of both the source and the foreground mass, such as pixelated images. When both the projected density of the lens and the image of the source are represented as pixelated images, the lensed image depends non-linearly on the foreground mass distribution. The resulting inference task then constitutes a high-dimensional non-linear inverse problem. Such problems are typically characterized by highly complex, multimodal posterior distributions and severe parameter degeneracies that often render traditional, as well as modern, diffusion-based sampling algorithms ineffective (see, e.g., Appendix E of \cite{Sharief_Zeghal_Barco_Lemos_Hezaveh_Perreault-Levasseur_2026}).

While methods leveraging flexible representations of the source \citep{Warren_Dye_2003, Suyu_Marshall_Hobson_Blandford_2006, Nightingale_Dye_2015} and foreground mass \citep{Koopmans_2005, Saha_Coles_Maccio_Williams_2006, Vegetti_Koopmans_Bolton_Treu_Gavazzi_2009} exist, they typically employ analytic prior distributions chosen for their simplicity. A previous work \citep{Adam_Perreault-Levasseur_Hezaveh_Welling_2023} discussed a method to obtain point estimates of the source and foreground mass as pixelated images which leveraged flexible data-driven priors, but sampling from the corresponding joint posterior distribution has remained an open problem.

\begin{figure*}[t]
    \centering
    \resizebox{0.8\textwidth}{!}{
        \input{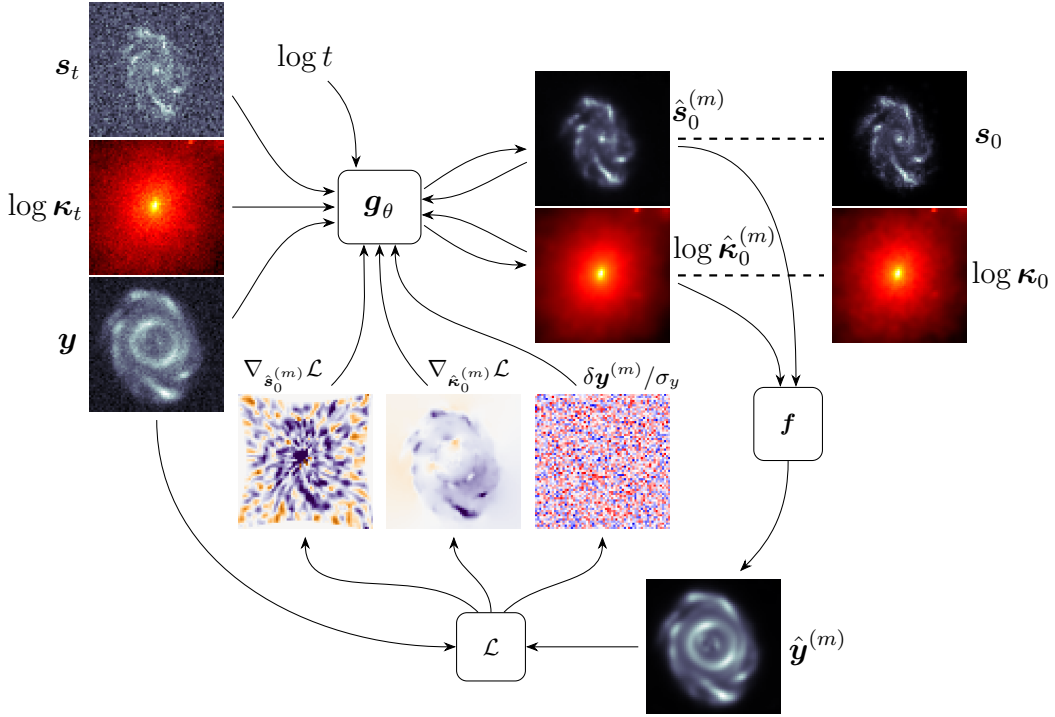}
    }
    \caption{Graphical representation of DiRIM (Diffusion Recurrent Inference Machine) for strong gravitational lensing. A neural network $\boldsymbol{g}_\theta$ is trained to denoise noisy source galaxy images $\boldsymbol{s}_t$ and noisy convergence map images $\log\boldsymbol{\kappa}_t$ given lens observations $\boldsymbol{y}$. It iteratively updates its estimate of the denoised source and convergence map $(\hat{\boldsymbol{s}}_0, \log\hat{\boldsymbol{\kappa}}_0)$ using the Recurrent Inference Machine (RIM) framework. The dashed lines represent the loss function Eq.\ref{eq:loss DiRIM} (with $\boldsymbol{x} \equiv (\boldsymbol{s},\log \boldsymbol{\kappa}))$, and the trained $\boldsymbol{g}_\theta$ is used to determine posterior scores $\nabla_{\boldsymbol{s}_t,\log\boldsymbol{\kappa}_t} \log p_t(\boldsymbol{s}_t,\log\boldsymbol{\kappa}_t|\boldsymbol{y})$ needed to obtain samples from the posterior $p(\boldsymbol{s}_t,\log\boldsymbol{\kappa}_t|\boldsymbol{y})$ via diffusion-based generative modeling.
    }
    \label{fig:DiRIM lensing}
\end{figure*}

In this work, we develop a Bayesian method to sample from the joint posterior of the source galaxy brightness and foreground mass distribution as pixelated images conditioned on observations. Our method, named Diffusion Recurrent Inference Machine (DiRIM), is based on diffusion-based generative models and RIMs. It works by using a RIM to determine scores of the posterior distribution and subsequently solving a reverse stochastic differential equation (SDE) to generate the posterior samples. This key idea is summarized in Figure \ref{fig:DiRIM lensing}.

In this paper, we train a DiRIM on realistic mock observations of strong gravitational lenses generated using source galaxy and foreground mass images taken from cosmological magneto-hydrodynamical simulations, and show that we can model the observations down to the noise level. This constitutes a first proof of concept regarding DiRIM's ability to solve high-dimensional non-linear inverse problems.

The outline of the paper is as follows. In Section \ref{sec:methods}, we provide an overview of diffusion-based generative models and RIMs, then describe our proposed DiRIM framework and the datasets we use in this work. In Section \ref{sec:results}, we present our main results and discuss their significance. Finally, in Section \ref{sec:conclusion}, we discuss our conclusions.

\section{Methods} \label{sec:methods}

\subsection{Background}

\subsubsection{Strong gravitational lensing simulations}

Strong gravitational lensing simulations are implemented using the thin screen approximation \citep{Meneghetti_2022} wherein light deflection is assumed to occur on a single plane called the lens plane. The lensed images are computed using ray tracing, with the lens plane coordinates $\boldsymbol{\theta} = \{\theta_x,\theta_y\}$ of the rays related to their source plane coordinates $\boldsymbol{\beta} = \{\beta_x,\beta_y\}$ by the lens equation
\begin{align}
    \boldsymbol{\beta}(\boldsymbol{\theta}) = \boldsymbol{\theta} - \boldsymbol{\alpha}(\boldsymbol{\theta}).
\end{align}
The reduced deflection angle $\boldsymbol{\alpha}(\boldsymbol{\theta})$ is calculated from the dimensionless projected surface density field $\kappa(\boldsymbol{\theta})$, also referred to as convergence map, as
\begin{align}
    \boldsymbol{\alpha}(\boldsymbol{\theta}) = \frac{1}{\pi} \int d^2\boldsymbol{\theta}' \kappa(\boldsymbol{\theta}')\frac{\boldsymbol{\theta}-\boldsymbol{\theta}'}{|\boldsymbol{\theta}-\boldsymbol{\theta}'|^2}.
\end{align}
In the simulated lensed images, the light intensity of a pixel at angular position $\boldsymbol{\theta}$ is given by the source intensity at the corresponding source plane angular position $\boldsymbol{\beta}(\boldsymbol{\theta})$. The lensed image is then convolved with a point spread function (PSF), and Gaussian additive instrument noise is applied to produce a simulated observation. This defines a forward model
\begin{align}
    \boldsymbol{y} = \boldsymbol{f}(\boldsymbol{s},\boldsymbol{\kappa}) + \boldsymbol{\eta} \label{eq:forward model},
\end{align}
where $\boldsymbol{s}$ is the source brightness distribution, $\boldsymbol{\kappa}$ is the convergence map, $\boldsymbol{f}$ is the forward operator consisting of ray tracing and PSF convolution, $\boldsymbol{\eta}$ is the additive observational noise and $\boldsymbol{y}$ is the simulated observation.

When $\boldsymbol{s}$ and $\boldsymbol{\kappa}$ are represented by pixelated images, the deflection angles and source brightnesses at the required angular positions are obtained using bilinear interpolation of the pixelated deflection angle images and pixelated source brightness images, respectively.

The forward operator $\boldsymbol{f}$ is implemented in \textsc{Caustics} \citep{Stone_Adam_Coogan_Yantovski-Barth_Filipp_Setiawan_Core_Legin_Wilson_Barco_et_al._2024}, a GPU-accelerated, automatically differentiable gravitational lensing simulator.

\subsubsection{Diffusion models} \label{sec:diffusion}

Diffusion-based generative models are a class of machine learning models that can generate samples from complex probability distributions $p(\boldsymbol{x})$. Here, $\boldsymbol{x} \in \mathbb{R}^{n}$ are n-dimensional vectors (they could be, for example, vectors containing the values of the n pixels of an image).

In their formulation as score-based models, diffusion models can be described using SDEs \citep{Song_Sohl-Dickstein_Kingma_Kumar_Ermon_Poole_2020}. In this description, a first SDE, referred to as the forward SDE
\begin{align}
    d\boldsymbol{x} = \boldsymbol{f}(\boldsymbol{x},t)dt + g(t)d\boldsymbol{w}(t), 
    \label{eq:forward SDE}
\end{align}
is defined, where $\boldsymbol{w}(t)$ is the standard Brownian motion, $\boldsymbol{
f}(\boldsymbol{x},t)$ is the drift coefficient, and $g(t)$ is the diffusion coefficient. The forward SDE is chosen so that when evolved over the time interval $t\in[0,T]$ (where $T$ is typically set to 1), it transports samples from $p(\boldsymbol{x})$ into samples from a Gaussian distribution\footnote{For the choice of SDE in practical implementations, the SDE transports samples from $p(\boldsymbol{x})$ into samples from a distribution that is only approximately Gaussian.}. Intuitively, this forward process can be thought of as the diffusion of ink molecules dissolving in a liquid, which gradually erases the structure of the initial physical state until a featureless, maximum-entropy distribution remains.

In order to generate samples from $p(\boldsymbol{x})$, one generates samples from the Gaussian distribution and transports them back in time from $t=T$ to $t=0$ by solving the so-called reverse-time SDE \citep{Anderson_1982}
\begin{align}
    d\boldsymbol{x} = [\boldsymbol{f}(\boldsymbol{x},t) - g(t)^2 \nabla_{\boldsymbol{x}} \log p_t(\boldsymbol{x})] dt + g(t)d\bar{\boldsymbol{w}}(t), \label{eq:reverse SDE}
\end{align}
where $\bar{\boldsymbol{w}}$ is the reverse-time standard Brownian motion and $p_t(\boldsymbol{x})$ is the probability density of $\boldsymbol{x}$ at time $t$. Doing so requires knowing the quantity $\nabla_{\boldsymbol{x}}\log p_t(\boldsymbol{x})$, known as the score of $p_t(\boldsymbol{x})$, for all $t \in [0,T]$. Intuitively, this reverse process can be thought of as rewinding the diffusion of the ink molecules. However, to successfully force the dispersed particles back into their original, structured configuration and reduce the system's entropy, an external guiding force is required; this is described by the score.

For non-analytical probability distributions $p(\boldsymbol{x})$, a closed-form expression for $\nabla_{\boldsymbol{x}}\log p_t(\boldsymbol{x})$ is unknown. It is therefore approximated using deep learning. Given a dataset $\{\boldsymbol{x}^{(i)}\}_{i=1}^{N}$ of samples $\boldsymbol{x}^{(i)} \sim p(\boldsymbol{x})$, one uses a neural network $\boldsymbol{s}_{\theta}(\boldsymbol{x}_t,t)$ with parameters $\theta$ trained by minimizing the denoising score-matching loss function \citep{Hyvrinen2005EstimationON, Vincent_2011, Song_Sohl-Dickstein_Kingma_Kumar_Ermon_Poole_2020}
\begin{multline}
    L_{\text{DSM}}=\mathbb{E}_{t\sim \mathcal{U}[0,T],\boldsymbol{x}_0\sim p(\boldsymbol{x}),\boldsymbol{x}_t\sim p(\boldsymbol{x}_t|\boldsymbol{x}_0)}\big\{ \\ \lambda(t)\|\boldsymbol{s}_\theta(\boldsymbol{x}_t,t)-\nabla_{\boldsymbol{x}_t}\log p(\boldsymbol{x}_t|\boldsymbol{x}_0)\|_2^2\big\}.
\end{multline}
It can be shown that in the limit of infinite data and model capacity, the set of parameters $\theta^\ast$ minimizing the loss function $L_{\text{DSM}}$ satisfy $s_{\theta^\ast}(\boldsymbol{x},t)$ = $\nabla_{\boldsymbol{x}}\log p_t(\boldsymbol{x})$ for almost all $\boldsymbol{x}$ and $t$ \citep{Song_Sohl-Dickstein_Kingma_Kumar_Ermon_Poole_2020}. Therefore, a trained $\boldsymbol{s}_\theta(\boldsymbol{x},t)$ can be used to solve the reverse SDE Eq.\ref{eq:reverse SDE}.

When a forward model mapping model parameters $\boldsymbol{x}$ to observations $\boldsymbol{y}$ (e.g., Eq.\ref{eq:forward model} for which $\boldsymbol{x}=\{\boldsymbol{s},\boldsymbol{\kappa}\}$) is defined, this diffusion framework can be extended to generate samples from the posterior distribution $p(\boldsymbol{x}|\boldsymbol{y})$ for any $\boldsymbol{y}$. This is done by solving the reverse-time SDE
\begin{align}
    d\boldsymbol{x} = [f(\boldsymbol{x},t) - g(t)^2 \nabla_{\boldsymbol{x}} \log p_t(\boldsymbol{x}|\boldsymbol{y})] dt + g(t)d\bar{\boldsymbol{w}}(t), \label{eq:reverse SDE posterior}    
\end{align}
which requires knowing the score $\nabla_{\boldsymbol{x}}\log p_t(\boldsymbol{x}|\boldsymbol{y})$. This score can be approximated using a neural network $\boldsymbol{s}_\theta(\boldsymbol{x},\boldsymbol{y},t)$, 
which has the observation $\boldsymbol{y}$ as an additional input, and by minimizing the conditional denoising score-matching loss function
\begin{multline}
    L_{\text{CDSM}}=\mathbb{E}_{t\sim \mathcal{U}[0,T],\boldsymbol{x}_0\sim p(\boldsymbol{x}), \boldsymbol{y}\sim p(\boldsymbol{y}|\boldsymbol{x}_0),\boldsymbol{x}_t\sim p(\boldsymbol{x}_t|\boldsymbol{x}_0)}\big\{ \\\lambda(t)\|\boldsymbol{s}_\theta(\boldsymbol{x}_t,\boldsymbol{y},t)-\nabla_{\boldsymbol{x}_t}\log p_t(\boldsymbol{x}_t|\boldsymbol{x}_0)\|_2^2\big\}. \label{eq:loss CDSM explicit}
\end{multline} 
One can show that the set of parameters $\theta^\ast$ minimizing $L_{\text{CDSM}}$ satisfy, in the limit of infinite data and model capacity, $\boldsymbol{s}_{\theta^\ast}(\boldsymbol{x},\boldsymbol{y},t)$ = $\nabla_{\boldsymbol{x}}\log p_t(\boldsymbol{x}|\boldsymbol{y})$ for almost all $\boldsymbol{x}$, $\boldsymbol{y}$, and $t$ \citep{Batzolis_Stanczuk_Schonlieb_Etmann_2021}.

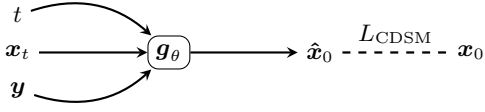
\begin{figure}
    \centering
    \begin{tikzpicture}[node distance=2cm]
    \node(xt) [] {$\boldsymbol{x}_t$};
    \node(t) [above of=xt, node distance=0.5cm] {$t$};
    \node(y) [below of=xt, node distance=0.5cm] {$\boldsymbol{y}$};
    
    \node (g) [draw, rounded corners, right of=xt, node distance=2cm] {$\boldsymbol{g}_{\theta}$};
    
    \node(xhat) [right of=g, node distance=2cm,
          yshift=0cm] {$\boldsymbol{\hat{x}}_0$};
    
    \node(x0) [right of=xhat, node distance=2cm] {$\boldsymbol{x}_0$};
    
    \draw [arrow] (t) to[bend left] (g);
    \draw [arrow] (xt) to[] (g);
    \draw [arrow] (y) to[bend right] (g);
    \draw [arrow] (g) to[] (xhat);
    
    \draw [dashed, line width=0.3mm] (xhat) -- (x0) node[midway, above] {$L_\text{CDSM}$};    
\end{tikzpicture}
    \caption{Graphical representation of conditional denoising score matching. A neural network $\boldsymbol{g}_\theta$ is trained to denoise noisy model parameters $\boldsymbol{x}_t$ generated by passing prior samples $\boldsymbol{x}_0$ through the forward SDE Eq.\ref{eq:forward SDE}. The neural network does so given $t$ and observations $\boldsymbol{y}$ generated from $\boldsymbol{x}_0$ by the forward model. It outputs an estimate $\hat{\boldsymbol{x}}_0$ of $\boldsymbol{x}_0$, and the dashed line represents the loss function Eq.\ref{eq:loss CDSM}.
    }
    \label{fig:cdsm}
\end{figure}

In practice, a popular choice of forward SDE is the variance exploding (VE) SDE \citep{Song_Sohl-Dickstein_Kingma_Kumar_Ermon_Poole_2020}
\begin{align}
    d\boldsymbol{x} = \sqrt{\frac{d[\sigma^2(t)]}{dt}}d\boldsymbol{w}, \label{eq:VE SDE}
\end{align}
where $\sigma(t)$ is a monotonically increasing function satisfying $\sigma(0) \approx 0$. In this case, the loss function $L_{\text{CDSM}}$ can be written as 
\begin{multline}
    L_{\text{CDSM}}=\mathbb{E}_{t\sim \mathcal{U}[0,T],\boldsymbol{x}_0\sim p(\boldsymbol{x}), \boldsymbol{y}\sim p(\boldsymbol{y}|\boldsymbol{x}_0),\boldsymbol{x}_t\sim p(\boldsymbol{x}_t|\boldsymbol{x}_0)}\big\{ \\ \frac{\lambda(t)}{\sigma^4(t)}\|\boldsymbol{g}_\theta(\boldsymbol{x}_t,\boldsymbol{y},t)-\boldsymbol{x}_0\|_2^2\big\}, \label{eq:loss CDSM}
\end{multline}
where $\boldsymbol{g}_\theta$ is related to $\boldsymbol{s}_\theta$ by 
\begin{align}
    \boldsymbol{s}_\theta = \frac{\boldsymbol{g}_\theta - \boldsymbol{x}_t}{\sigma^2(t)}, \label{eq:score denoising relation}
\end{align}
(see Appendix \ref{app:sm denoising equivalence} for details). Eq.\ref{eq:loss CDSM} implies that $\boldsymbol{g}_\theta(\boldsymbol{x}_t,\boldsymbol{y},t)$ can be trained as a denoising model recovering $\boldsymbol{x}_0$ given $\boldsymbol{x}_t$. This is represented graphically in Figure \ref{fig:cdsm}. 

\subsubsection{Recurrent inference machines} \label{sec:RIMs}

Recurrent Inference Machines (RIMs) are a machine learning framework for solving inverse problems. The task in inverse problems is to reconstruct a signal $\boldsymbol{x}$ from a noisy observation $\boldsymbol{y}$ generated via a forward process
\begin{align}
    \boldsymbol{y} = \boldsymbol{f}(\boldsymbol{x}) + \boldsymbol{\eta},
\end{align}
where $\boldsymbol{f}$ is the forward operator and $\boldsymbol{\eta}$ is additive observational noise. 

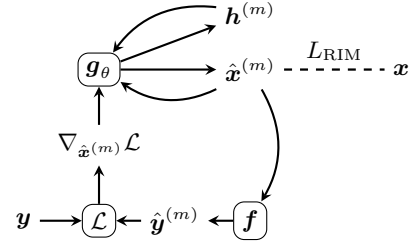
\begin{figure}
    \centering
      \begin{tikzpicture}[node distance=2cm]

    \node (g) [draw, rounded corners] {$\boldsymbol{g}_{\theta}$};

    \node(xhat) [right of=g, node distance=2cm,
          yshift=0cm] {$\hat{\boldsymbol{x}}^{(m)}$};

    \node(x0) [right of=xhat, node distance=2cm] {$\boldsymbol{x}$};

    \node(f) [draw, rounded corners, below of=xhat, node distance=2.0cm] {$\boldsymbol{f}$};

    \node(yhat) [left of=f, node distance=1cm] {$\hat{\boldsymbol{y}}^{(m)}$};

    \node (likelihood) [draw, rounded corners, left of=yhat, node distance=1cm] {$\mathcal{L}$};

    \node(y) [left of=likelihood, node distance=1cm] {$\boldsymbol{y}$};
    
    \node(nabla) [above of=likelihood, node distance=1.0cm, xshift=0cm]{$\nabla_{\hat{\boldsymbol{x}}^{(m)}}\mathcal{L}$};

    \node(h) [above of=xhat, node distance=0.75cm] {$\boldsymbol{h}^{(m)}$};

    \draw [arrow] (g) to[] (xhat);
    \draw [arrow] (xhat) to[bend left] (g);
    \draw [dashed, line width=0.3mm] (xhat) -- (x0) node[midway, above] {$L_\text{RIM}$};
    \draw [arrow] (xhat) to[bend left] (f);
    \draw [arrow] (f) to[] (yhat);
    \draw [arrow] (yhat) to[] (likelihood);
    \draw [arrow] (y) to[] (likelihood);
    \draw [arrow] (likelihood) to[] (nabla);
    \draw [arrow] (nabla) to[] (g);
    \draw [arrow] (g) to[] (h);
    \draw [arrow] (h) to[bend right] (g);    
  \end{tikzpicture}
    \caption{Graphical representation of RIMs. A neural network $\boldsymbol{g}_\theta$ is trained to solve inverse problems by producing point estimates $\hat{\boldsymbol{x}}$ of the model parameters $\boldsymbol{x}$ that produced an observation $\boldsymbol{y}$ through the forward model $\boldsymbol{f}$. The neural network iteratively updates its predictions $\hat{\boldsymbol{x}}$ using gradients of the negative log likelihood function $\mathcal{L} \equiv -\log p(\boldsymbol{y}|\hat{\boldsymbol{x}})$. The dashed line represents the loss function Eq.\ref{eq:loss RIM}.
    }
    \label{fig:RIM}
\end{figure}

In their original implementation \citep{Putzky_Welling_2017}, RIMs aim to determine the maximum a posteriori (MAP) solution of the inverse problem defined by
\begin{align}
    \boldsymbol{x}_{\text{MAP}} = \max_{\boldsymbol{x}} \log p(\boldsymbol{x}|\boldsymbol{y}).
\end{align}
Given the gradient of the negative log likelihood function, $\nabla_{\boldsymbol{x}}\mathcal{L}$, where $\mathcal{L}\equiv-\log p(\boldsymbol{y}|\boldsymbol{x})$, RIMs determine $\boldsymbol{x}_\text{MAP}$ within a recurrent neural network framework. In detail, $\boldsymbol{x}_\text{MAP}$ is approximated recursively via the update equations 
\begin{align}
    \hat{\boldsymbol{x}}^{(m+1)} &= \hat{\boldsymbol{x}}^{(m)} + \boldsymbol{g}_\theta(\hat{\boldsymbol{x}}^{(m)}, \boldsymbol{h}^{(m)}, \nabla_{\hat{\boldsymbol{x}}^{(m)}}\mathcal{L}) \nonumber \\
    \boldsymbol{h}^{(m+1)} &= \boldsymbol{g}_\theta^\ast(\hat{\boldsymbol{x}}^{(m)}, \boldsymbol{h}^{(m)}, \nabla_{\hat{\boldsymbol{x}}^{(m)}}\mathcal{L}),
\end{align}
where $\boldsymbol{g}_\theta$ is a neural network with parameters $\theta$ and $\boldsymbol{g}^\ast_\theta$ is the update model for the hidden state variables $\boldsymbol{h}$, which constitute the RIM's memory and allow it to track its progression through the refinement iterations $m \in \{1,\cdots,M\}$. This is represented graphically in Figure \ref{fig:RIM}.

The RIM parameters $\theta$ are determined by minimizing the mean squared error loss function
\begin{align}
    L_{\text{RIM}} = \mathbb{E}_{\boldsymbol{x}\sim p(\boldsymbol{x}),\boldsymbol{y}\sim p(\boldsymbol{y}|\boldsymbol{x})}\bigg\{\sum_{m=1}^{M} w_m \|\hat{\boldsymbol{x}}^{(m)}-\boldsymbol{x}\|_2^2\bigg\}, \label{eq:loss RIM}
\end{align}
where $\hat{\boldsymbol{x}}^{(m)}$ is the prediction of the RIM at refinement iteration $m$, $\boldsymbol{x}$ is the true parameters and $w_m$ is a set of RIM iteration loss weights. Since in Eq.\ref{eq:loss RIM}, the expectation is taken over $\boldsymbol{x}\sim p(\boldsymbol{x})$, after training the network weights encode the prior distribution implicitly and evaluating the RIM at test time will optimize the likelihood guided by the expectation over this encoded prior, making the output converge to the MAP.

\subsection{DiRIM} \label{sec:dirim}

This section introduces our DiRIM framework and how we apply it to the problem of strong gravitational lensing.

\subsubsection{DiRIM framework}

The purpose of DiRIM is to combine the ability of diffusion models to sample from complex high-dimensional distributions and the ability of RIMs to solve non-linear inverse problems into a single framework that can sample from complex, possibly highly degenerate, high-dimensional posterior distributions $p(\boldsymbol{x}|\boldsymbol{y})$ for which the forward model $\boldsymbol{y} = \boldsymbol{f}(\boldsymbol{x}) + \boldsymbol{\eta}$ is non-linear.  

\begin{figure}
    \centering
      \begin{tikzpicture}[node distance=2cm]

    \node (g) [draw, rounded corners] {$\boldsymbol{g}_{\theta}$};

    \node (xt) [left of=g, node distance=1.5cm] {$\boldsymbol{x}_t$};
    
    \node(y) [below of=xt, node distance=0.5cm] {$\boldsymbol{y}$};

    \node(t) [above of=xt, node distance=0.5cm] {$t$};

    \node(xhat) [right of=g, node distance=2cm,
          yshift=0cm] {$\hat{\boldsymbol{x}}_0^{(m)}$};

    \node(x0) [right of=xhat, node distance=2cm] {$\boldsymbol{x}_0$};

    \node(f) [draw, rounded corners, below of=xhat, node distance=2.0cm] {$\boldsymbol{f}$};

    \node(yhat) [left of=f, node distance=1cm] {$\hat{\boldsymbol{y}}^{(m)}$};

    \node (likelihood) [draw, rounded corners, left of=yhat, node distance=1cm] {$\mathcal{L}$};
    
    \node(nabla) [above of=likelihood, node distance=1.0cm, xshift=0cm]{$\nabla_{\hat{\boldsymbol{x}}_0^{(m)}} \mathcal{L}$};

    \node(h) [above of=xhat, node distance=0.75cm] {$\boldsymbol{h}^{(m)}$};

    \draw[arrow] (t) to[bend left] (g);
    \draw[arrow] (xt) to[] (g);
    \draw [arrow] (g) to[] (xhat);
    \draw [arrow] (xhat) to[bend left] (g);
    \draw [arrow] (y) to[bend right] (g);
    \draw [dashed, line width=0.3mm] (xhat) -- (x0) node[midway, above] {$L_\text{DiRIM}$};
    \draw [arrow] (xhat) to[bend left] (f);
    \draw [arrow] (f) to[] (yhat);
    \draw [arrow] (yhat) to[] (likelihood);
    \draw [arrow] (y) to[bend right] (likelihood);
    \draw [arrow] (likelihood) to[] (nabla);
    \draw [arrow] (nabla) to[] (g);
    \draw [arrow] (g) to[] (h);
    \draw [arrow] (h) to[bend right] (g);    
  \end{tikzpicture}
    \caption{Graphical representation of DiRIM. A neural network $\boldsymbol{g}_\theta$ is trained to denoise noisy model parameters $\boldsymbol{x}_t$ given $t$ and observations $\boldsymbol{y}$. It does so by iteratively updating its prediction $\hat{\boldsymbol{x}}_0$ using the RIM framework. The dashed line represents the loss function Eq.\ref{eq:loss DiRIM}. This computation graph corresponds to a superimposition of the computation graphs in Figures \ref{fig:cdsm} and \ref{fig:RIM}. 
    }
    \label{fig:DiRIM}
\end{figure}
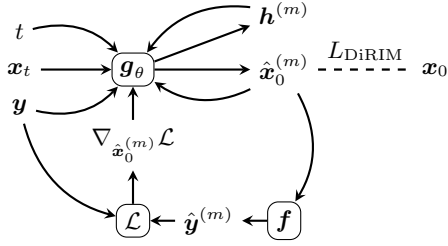

Intuitively, the DiRIM framework amounts to using a conditional diffusion model to determine the posterior scores necessary to sample the posterior distribution, with the added feature that these posterior scores are refined recursively using the recurrent inference machine framework. 
This way, when solving a reverse SDE to generate a posterior sample, the posterior scores used at every step are more accurate, leading to a better calibrated sample. Concretely, this is done by creating a conditional diffusion model $\boldsymbol{g}_\theta$ (see Section \ref{sec:diffusion}) that is itself a RIM (see Section \ref{sec:RIMs}), as shown graphically in Figure \ref{fig:DiRIM}.

The loss function used to train the DiRIM essentially remains the conditional denoising score matching loss used to train conditional diffusion models. Specifically, we use the loss function
\begin{multline}
    L_{\text{DiRIM}} =\mathbb{E}_{t\sim \mathcal{U}[0,T],\boldsymbol{x}_0\sim p(\boldsymbol{x}),\boldsymbol{y}\sim p(\boldsymbol{y}|\boldsymbol{x}_0),\boldsymbol{x}_t\sim p(\boldsymbol{x}_t|\boldsymbol{x}_0)}\bigg\{\\ W(t) \sum_{m=1}^{M} w_m \|\hat{\boldsymbol{x}}_0^{(m)}-\boldsymbol{x}_0\|_2^2\bigg\}, \label{eq:loss DiRIM}
\end{multline}

with our choice of time weight function $W(t)$ detailed in Appendix \ref{app:lambda}. For the choice of RIM iteration loss weights
\begin{align}
\begin{cases}
    w_m = 1 & m = M\\
    w_m=0 & \text{else},
\end{cases} \label{eq:weights w_m}
\end{align}
the DiRIM loss function $L_{\text{DiRIM}}$ in Eq.\ref{eq:loss DiRIM} has the same minimizer as $L_{\text{CDSM}}$ in Eq.\ref{eq:loss CDSM} (see Appendix \ref{app:sm DiRIM and CDSM loss equivalence} for details). The consequence of this is that one may train the DiRIM as a denoising model via the loss function $L_{\text{DiRIM}}$ and subsequently (following Eq.\ref{eq:score denoising relation}) obtain posterior scores as
\begin{align}
    \nabla_{\boldsymbol{x}_t}\log p_t(\boldsymbol{x}_t|\boldsymbol{y}) = \frac{\boldsymbol{x}_0^{(M)}-\boldsymbol{x}_t}{\sigma^2(t)}. \label{eq:posterior score DiRIM VE SDE}
\end{align}
This expression is exact in the limit of infinite data and model capacity, and assumes usage of the VE SDE. We give the equivalent of Eq.\ref{eq:posterior score DiRIM VE SDE} for the variance preserving (VP) SDE \citep{Song_Sohl-Dickstein_Kingma_Kumar_Ermon_Poole_2020} in Appendix \ref{app:DiRIM VP}.

To generate samples from the posterior $p(\boldsymbol{x}|\boldsymbol{y})$, we proceed as described in Section \ref{sec:diffusion}, namely, we solve the reverse SDE Eq.\ref{eq:reverse SDE posterior} numerically using the posterior scores obtained from Eq.\ref{eq:posterior score DiRIM VE SDE} if using the VE SDE or Eq.\ref{eq:posterior score DiRIM VP SDE} if using the VP SDE. Our choice of SDE and our numerical approach to solving the reverse SDE are detailed in Appendix \ref{app:solving SDE}.

\subsubsection{DiRIM for gravitational lensing}

We apply the DiRIM framework to gravitational lensing with the goal of sampling from the posterior $p(\boldsymbol{s},\boldsymbol{\kappa}|\boldsymbol{y})$. For this purpose, we define $\boldsymbol{x}$ as the pair $(\boldsymbol{s},\log\boldsymbol{\kappa})$, comprising a source image and a convergence map image. We work with the log convergence $\log \boldsymbol{\kappa}$ rather than the convergence $\boldsymbol{\kappa}$ because we found it to perform better empirically. Gradients of the likelihood function, however, are still computed with respect to the convergence map itself. The full computation graph is shown in Figure \ref{fig:DiRIM lensing}.

Given that the noise $\boldsymbol{\eta}$ in the simulated observations is Gaussian ($\boldsymbol{\eta} \sim \mathcal{N}(0,\sigma_y^2 \boldsymbol{I})$), the likelihood function is also Gaussian and given by
\begin{align}
p(\boldsymbol{y}|\boldsymbol{s},\boldsymbol{\kappa)} = \mathcal{N}(\boldsymbol{f}(\boldsymbol{s},\boldsymbol{\kappa}),\sigma_y^2 \boldsymbol{I}).
\end{align}
Moreover, the likelihood function is differentiable as the forward operator $\boldsymbol{f}$ is implemented in \textsc{Caustics}, enabling the computation of the likelihood gradients $\nabla_{\boldsymbol{s}}\mathcal{L}$ and $\nabla_{\boldsymbol{\kappa}}\mathcal{L}$.

As a modification to the framework described in the previous section, we compute, at each DiRIM iteration $m$, the normalized lensed image residuals 
\begin{align}
    \frac{\delta \boldsymbol{y}^{(m)}}{\sigma_y} \equiv \frac{\boldsymbol{y}-\hat{\boldsymbol{y}}^{(m)}}{\sigma_y},
\end{align} and make $\boldsymbol{g}_\theta$ take $\delta\boldsymbol{y}^{(m)}/\sigma_y$ as input in addition to the likelihood gradients $\nabla_{\hat{\boldsymbol{s}}_0^{(m)}}\mathcal{L}$ and $\nabla_{\hat{\boldsymbol{\kappa}}_0^{(m)}}\mathcal{L}$. Additionally, we give $\log t$ as input to $\boldsymbol{g}_\theta$ as opposed to $t$. Finally, our experiments revealed no benefit to using a hidden state $\boldsymbol{h}$ across RIM iterations and therefore we do not include any. These adaptations are featured in Figure \ref{fig:DiRIM lensing}.

In this work, the iteratively refined estimates of denoised sources and convergence maps ($\boldsymbol{s}_0^{(m)}$ and $\log\boldsymbol{\kappa}_0^{(m)}$) are initialized as
\begin{align}
    \hat{\boldsymbol{s}}^{(0)}_0 &= \boldsymbol{s}_t \nonumber \\
    \log\hat{\boldsymbol{\kappa}}^{(0)}_0 &= \log\boldsymbol{\kappa}_t,
\end{align}
which takes advantage of the fact that for $t \approx 0$, $\boldsymbol{s}_t$ and $\log\boldsymbol{\kappa}_t$ are close to $\boldsymbol{s}_0$ and $\log\boldsymbol{\kappa}_0$.

\begin{table}
    \centering
    \begin{tabular}{c||c||c}
        \hline \hline
         Component & Parameter & Prior\\
         \hline
         EPL main halo &$x_l ['']$ & $\mathcal{U}[-0.12,0.12]$\\
         &$y_l ['']$ & $\mathcal{U}[-0.12,0.12]$\\
         &$q$ & $\mathcal{U}[0.7,1]$\\
         &$\phi$ & $\mathcal{U}[0,\pi]$\\
         &$R_{E} ['']$ & $\mathcal{U}[1,2]$\\
         &$\tau$ & $\mathcal{U}[0.75,1.25]$\\
         \hline
         Multipoles &$a_3 [('')^{-1}]$& $\mathcal{U}[0,0.05]$\\
         &$\theta_3$& $\mathcal{U}[0,2\pi/3]$\\
         &$a_4 [('')^{-1}]$ & $\mathcal{U}[0,0.05]$\\
         &$\theta_4$& $\mathcal{U}[0,\pi/2]$\\
         \hline
         NFW subhalo &$r_{\text{sub}} ['']$ & $\mathcal{U}[1.44,2.4]$\\
         &$\theta_{\text{sub}}$ & $\mathcal{U}[0,2\pi]$\\         
         &$\log_{10} M_{\text{sub}} [M_\odot]$ & $\mathcal{U}[10,11]$\\
         &$c_{\text{sub}}$ & $\mathcal{U}[50,100]$\\
    \end{tabular}
    \caption{Priors for all parameters used to create the analytic profile convergence maps.}
    \label{tab:priors}
\end{table}

In this work, the DiRIM's neural network $\boldsymbol{g}_\theta$ is a U-Net \citep{Ronneberger_Fischer_Brox_2015}. This is a common choice for diffusion models (see, e.g., \cite{Ho_Jain_Abbeel_2020, Song_Sohl-Dickstein_Kingma_Kumar_Ermon_Poole_2020}). 
We use the U-Net architecture discussed in \citet{Karras_Aittala_Aila_Laine_2022} which features residual blocks, skip connections, and the use of self-attention layers. The images input to $\boldsymbol{g}_\theta$ (namely, $\boldsymbol{s}_t,\log\boldsymbol{\kappa}_t,\boldsymbol{y},\nabla_{\hat{\boldsymbol{s}}_0^{(m)}}\mathcal{L},\nabla_{\hat{\boldsymbol{\kappa}}_0^{(m)}}\mathcal{L},\delta \boldsymbol{y}^{(m)}/\sigma_y,\hat{\boldsymbol{s}}_0^{(m)},\log \hat{\boldsymbol{\kappa}}_0^{(m)}$) are concatenated channel-wise, and the two output images ($\hat{\boldsymbol{s}}_0^{(m+1)},\log\hat{\boldsymbol{\kappa}}_0^{(m+1)}$) are produced as two channels.

Additional details regarding the neural network architecture and training procedure are discussed in Appendix \ref{app:unet}.

\begin{figure}[t]
    \centering
    \includegraphics[width=1\linewidth]{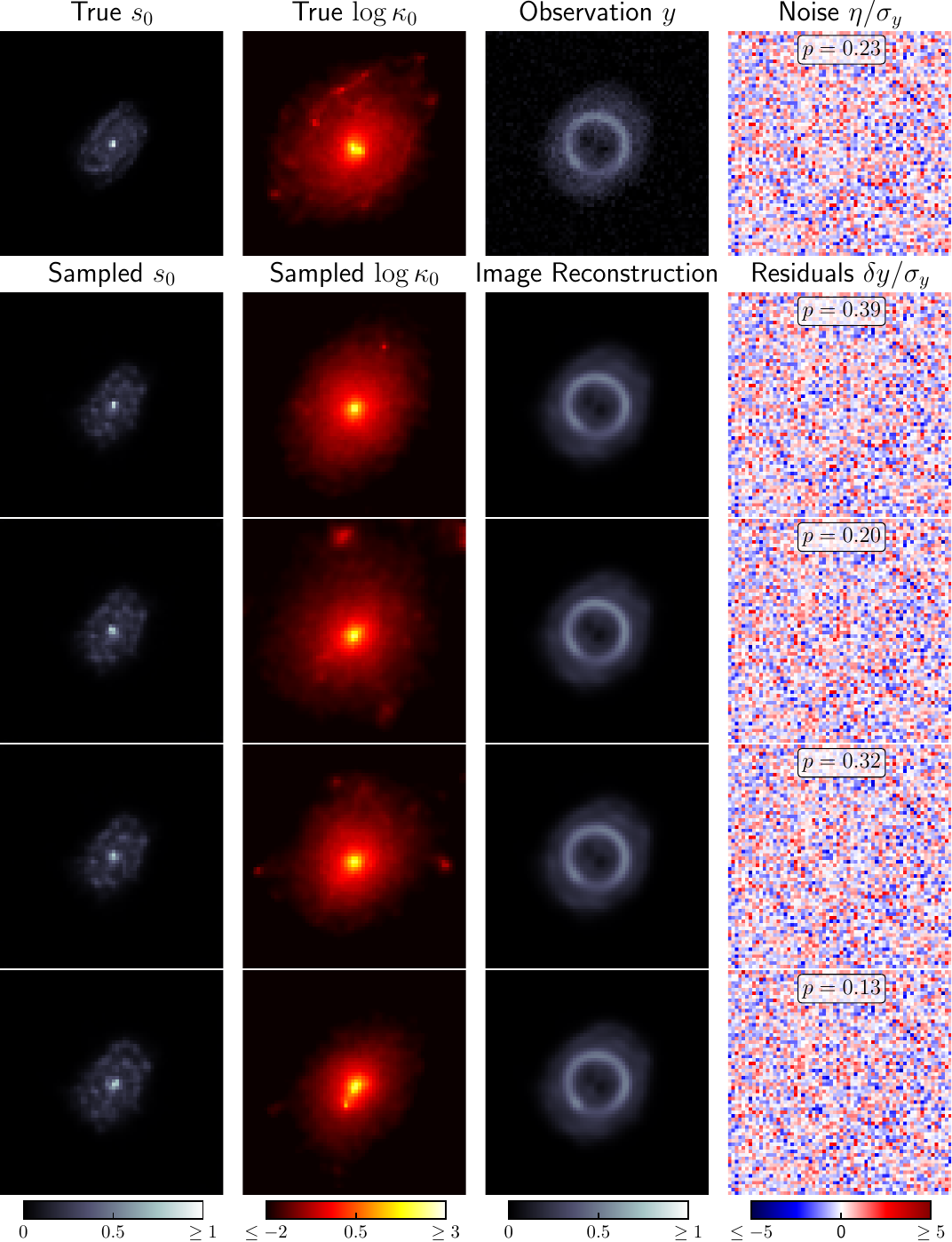}
    \caption{Lens model for a mock observation from the test set. This example features a convergence map with a single main deflector. We show the true source image, the true convergence map, the mock observation as well as four joint source and convergence map samples, with the reconstructed lensed image and residuals for each.}
    \label{fig:simulated convergence reconstruction simple}
\end{figure}

\subsection{Data}

\subsubsection{Source images} \label{subsubsec:source data}

In our experiments, the background source brightness distributions are taken from the SKIRT-TNG dataset \citep{Bottrell_Yesuf_Popping_Omori_Tang_Ding_Pillepich_Nelson_Eisert_Gao_et_al._2023}, which uses dust radiative transfer post-processing with SKIRT \citep{Camps_Baes_2020} to make a collection of images of simulated galaxies from the IllustrisTNG cosmological magneto-hydrodynamics simulations \citep{Nelson_Springel_Pillepich_Rodriguez-Gomez_Torrey_Genel_Vogelsberger_Pakmor_Marinacci_Weinberger_et_al._2019}. We use cutouts with a field of view (FOV) of 51.2 kiloparsec (kpc) from the {\it i}-band frames at $z=0$, downsampled to $64 \times 64$ pixels, to create a dataset of 42183 images. The images are normalized such that their brightest pixel is uniformly distributed between 0.9 and 1 in flux units of $\mu \text{Jy sr}^{-1}$. 

\subsubsection{Convergence map images} \label{subsubsec:convergence data}

To run our experiments, we create two separate convergence maps datasets. The first consists of simulations of dark matter and baryonic matter taken from the IllustrisTNG cosmological magneto-hydrodynamics simulations \citep{Nelson_Springel_Pillepich_Rodriguez-Gomez_Torrey_Genel_Vogelsberger_Pakmor_Marinacci_Weinberger_et_al._2019}. Halos with a dark matter mass of at least $9 \times 10^{11} M_\odot$ are selected, and smoothed convergence maps are created with an adaptive Gaussian kernel. For more details, see \citep{Adam_Perreault-Levasseur_Hezaveh_Welling_2023}, which used the same convergence maps. We select 42183 of those convergence maps.

\begin{figure}[t]
    \centering
    \includegraphics[width=1\linewidth]{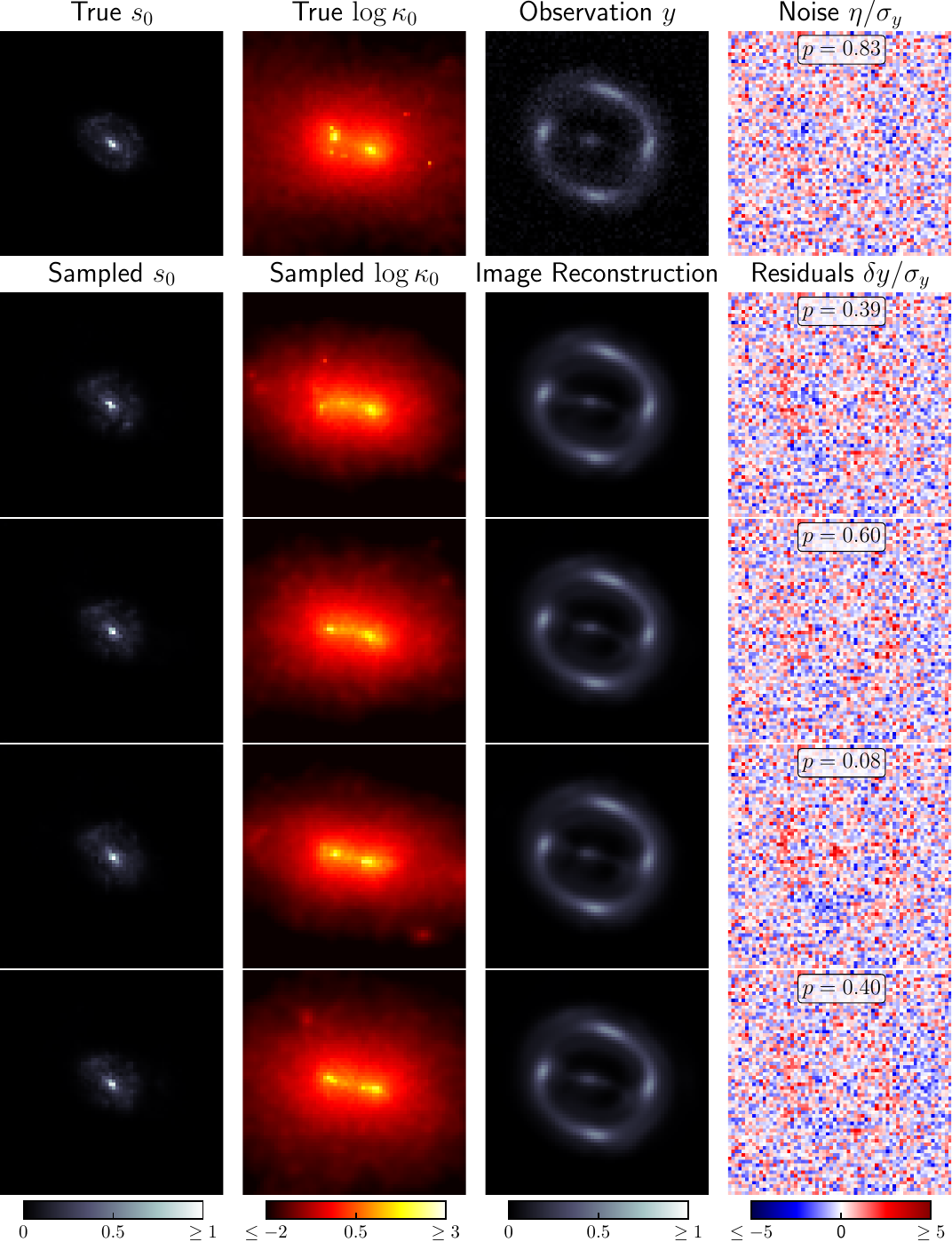}
    \caption{Same as Figure \ref{fig:simulated convergence reconstruction simple} for a different mock observation for which the convergence map contains two main deflectors, making it more complex.}
    \label{fig:simulated convergence reconstruction complex}
\end{figure}

\begin{figure}[t]
    \centering
    \includegraphics[width=1\linewidth]{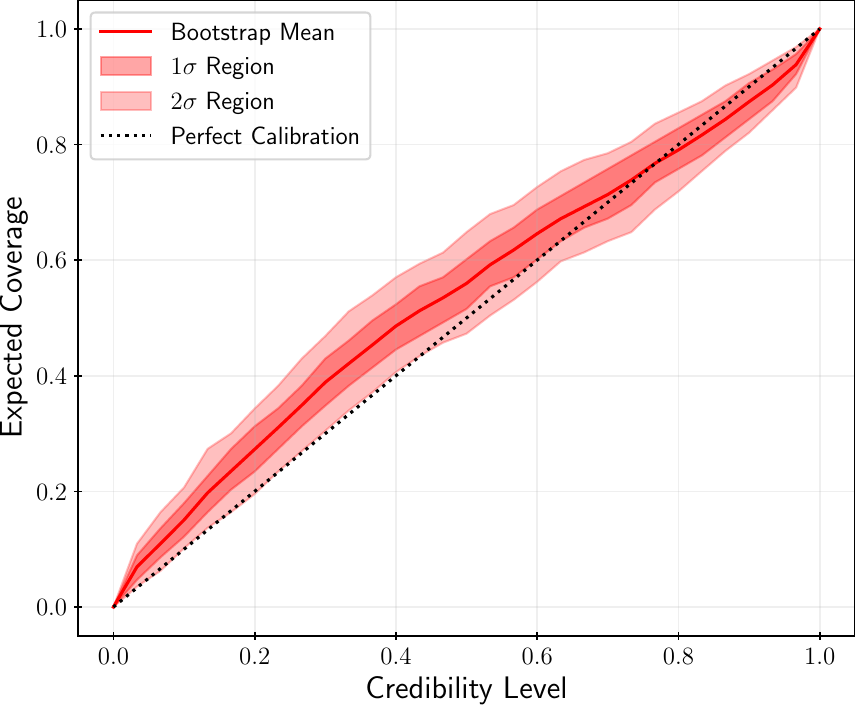}
    \caption{Tests of Accuracy with Random Points (TARP) plot computed on the test set. TARP is a necessary and sufficient condition for a posterior estimator to be unbiased. The plot demonstrates the near-perfect calibration of our posterior samples in pixel-space across the test set.}
    \label{fig:simulated convergence tarp}
\end{figure}

The second convergence map dataset consists of analytic profiles rendered on a pixelated grid. The profiles comprise an elliptical power law (EPL) \citep{Barkana_1998} main halo with convergence
\begin{align}
    \kappa_{\text{EPL}}(\theta_x,\theta_y) &= \frac{2-\tau}{2}\Bigg(\frac{R_E \sqrt{q}}{\sqrt{q^2 \theta_x^2 +\theta_y^2}}\Bigg)^\tau \label{eq:convergence EPL}
\end{align}
parameterized by the Einstein radius $R_E$, the density slope $\tau$  and the axis ratio $q$. In addition, the EPL component is rotated by an angle $\phi$. The EPL main halo is augmented with $m=3$ and $m=4$ multipoles with convergence
\begin{align}
    \kappa_{\text{multipoles}}(\theta_x,\theta_y) &= \sum_{m \in \{3,4\}} \frac{a_m}{2 r}\cos(m(\psi-\theta_m)), \label{eq:convergence multipoles}
\end{align}
where $r =\sqrt{\theta_x^2+\theta_y^2}$ and $\psi = \arctan(\theta_y/\theta_x)$. The multipoles are parameterized by the amplitudes $a_m$ and the orientations $\theta_m$. In addition, a Navarro-Frenk-White (NFW) \citep{Navarro_Frenk_White_1997} subhalo is added to $50 \%$ of the convergence maps, randomly. The convergence of the subhalos is the projection of the 3D mass density profile
\begin{align}
    \rho(r) = \frac{\rho_s}{(r/r_s)(1+r/r_s)^2}.
\end{align}
We parametrize the NFW subhalos by their mass $M_{\text{sub}}$ and concentration parameter $c_{\text{sub}}$. The equations relating $\rho_s$ and $r_s$ to $M_{\text{sub}}$ and $c_{\text{sub}}$ are given in \citep{Meneghetti_2022} (with $M_{\text{sub}} $ denoted by $M_{200}$). Finally, the EPL, external shear and multipole components are given common center angular coordinates parametrized by $(\boldsymbol{x}_0,\boldsymbol{y}_0)$, while the center of the subhalo is parametrized by the radial position parameters $r_{\text{sub}}$, $\theta_{\text{sub}}$. This corresponds to a total of 14 parameters, and a dataset of 42183 convergence maps is created by sampling these parameters according to the prior distribution in Table \ref{tab:priors}. The convergence maps are rendered on a $128 \times 128$ grid and subsequently downsampled to $64 \times 64$ by average pooling.

\subsubsection{Simulated observations}

The sets of 42183 background source images and convergence maps are each split into a training set (85\%), a validation set (10\%) and a test set (5\%).

\begin{figure}[t]
    \centering
    \includegraphics[width=1\linewidth]{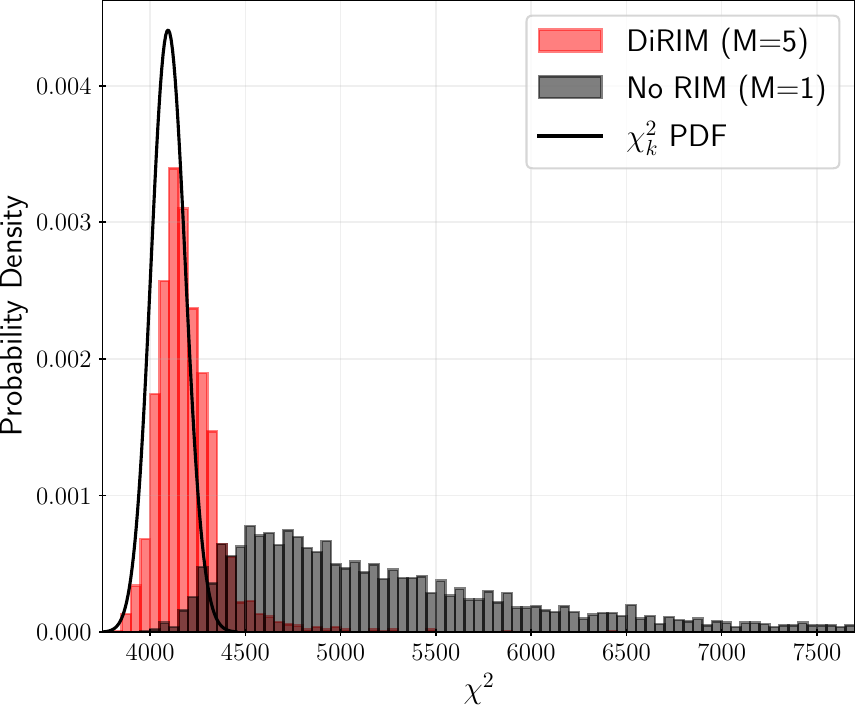}
    \caption{Distribution of $\chi^2$ values computed using simulated observations from the test set and the image reconstructions obtained from posterior samples. The $\chi^2$ value is computed for one posterior sample for each observation in the test set, and the resulting distribution is compared to the $\chi^2_k$ probability density function with $k=64^2$. The red histogram is obtained using our methods while the black histogram is obtained by removing the RIM from our methods, mimicking the behavior of baseline conditional denoising score matching.}
    \label{fig:simulated convergence chi2}
\end{figure}

\begin{figure*}[t]
    \centering
    \includegraphics[width=0.75\linewidth]{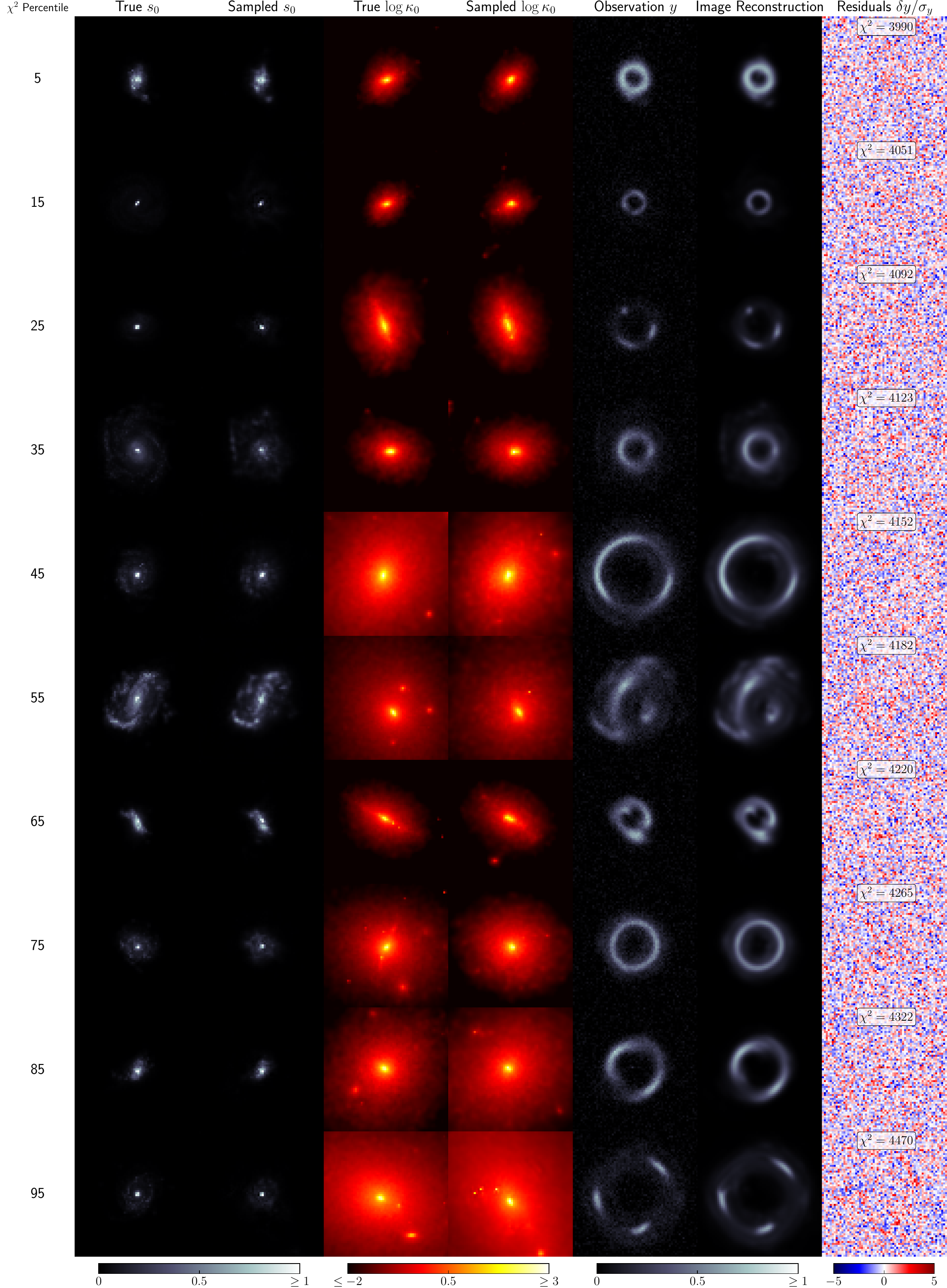}
    \caption{Posterior samples for mock observations from the test set, selected and ordered based on their reconstructed image $\chi^2$ percentile ranks. The samples are ordered from top to bottom in order of lowest $\chi^2$ to highest $\chi^2$.}
    \label{fig:simulated convergence chi2 percentile}
\end{figure*}

\begin{figure*}[t]
    \centering
    \includegraphics[width=0.7\linewidth]{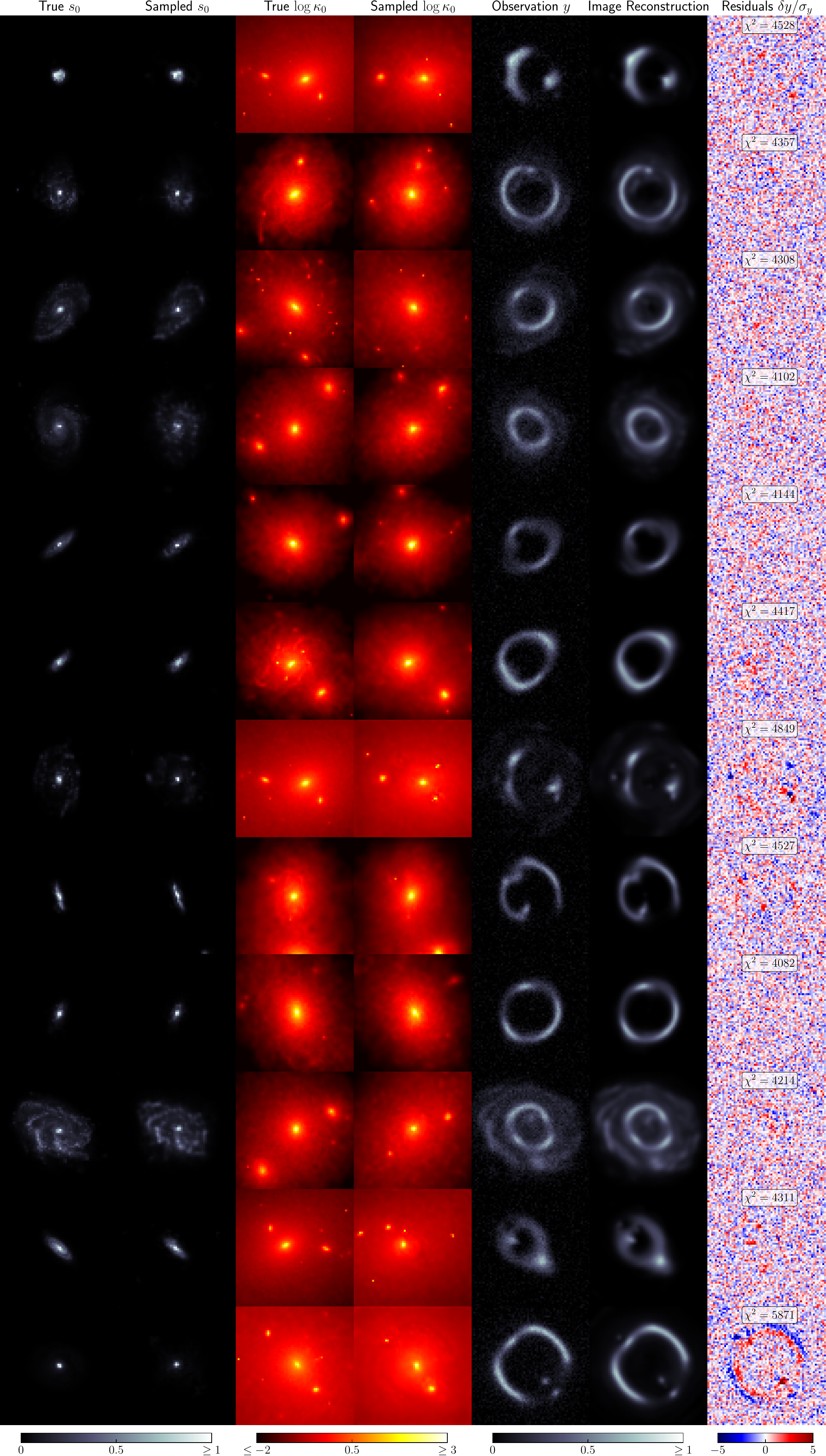}
    \caption{Posterior samples for mock observations from the test set, selected on the basis that their convergence map is highly complex. For each observation, we show one joint source and convergence map sample.}
    \label{fig:simulated convergence collection complex}
\end{figure*}

To create the simulated observations, the lens and source redshifts are fixed at $z_l=0.5$ and $z_s=1$, respectively, and a flat $\Lambda$CDM cosmology with $H_0=67.66 \text{ km}\text{ s}^{-1} \text{ Mpc}^{-1}$ and $\Omega_m=0.3097$ is assumed. A Gaussian PSF with a standard deviation of $0.12''$ is used. The lensed images are rendered at a size of $64 \times 64$ with a field of view of $7.68''$. We add Gaussian additive noise $\boldsymbol{\eta} \sim \mathcal{N}(0,\sigma_y^2 \boldsymbol{I})$ with $\sigma_y=0.03$ to the observations, resulting in a peak SNR uniformly distributed between $0.9/0.03=30$ and $1/0.03=33.33$.  

\section{Results and discussion} \label{sec:results}

Using the SKIRT-TNG source images, we train two separate DiRIM models; one on the set of simulated convergence maps and one on the set of analytic convergence maps. This section details our results for both models separately.

\subsection{Simulated convergence model}

This section discusses our results on tests done with the DiRIM model trained on the set of simulated convergence maps. The purpose of these tests is to assess the ability of our methods to model realistic gravitational lensing systems.

\begin{figure}[t]
    \centering
    \includegraphics[width=1\linewidth]{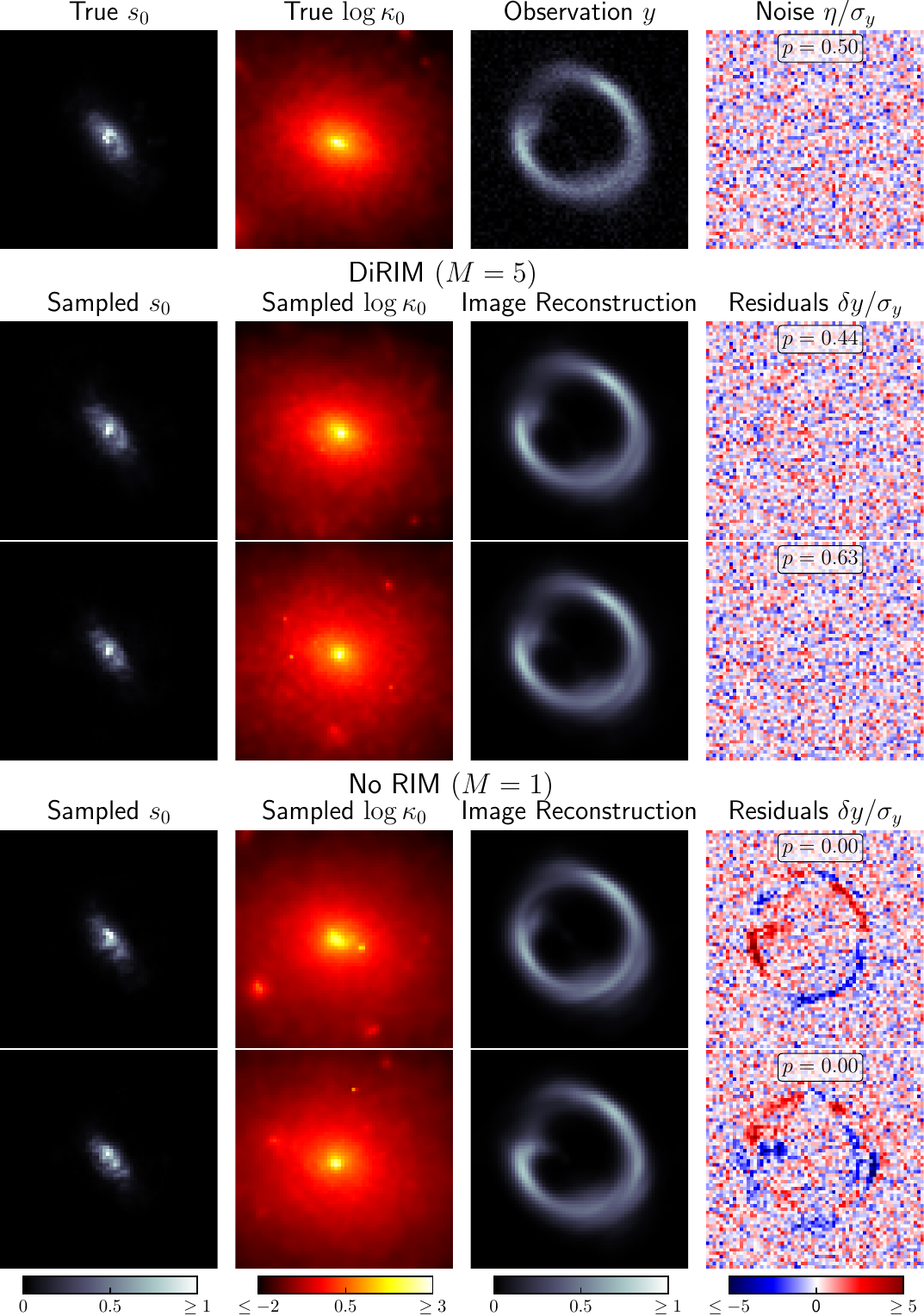}
    \caption{Comparison of posterior samples for an observation from the test set with the number of RIM iterations varied between $M=5$ and $M=1$. The latter disables the iterative refinement of posterior scores via the RIM framework, resulting in the inference being biased. This demonstrates that the RIM is an essential component of our methods.}
    \label{fig:simulated convergence comparison no rim}
\end{figure}

 \begin{figure}[t]
    \centering
    \includegraphics[width=1\linewidth]{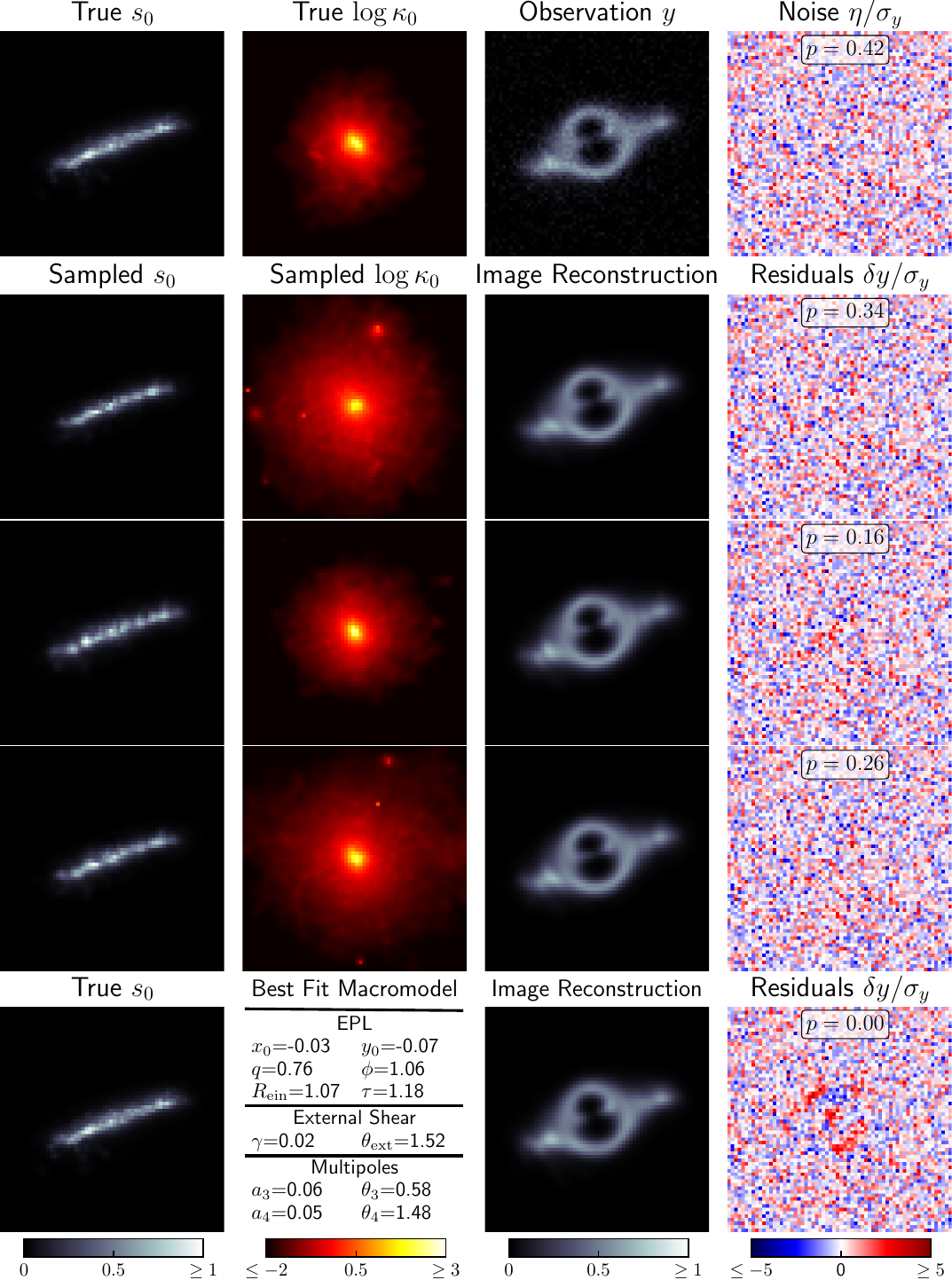}
    \caption{Comparison between our methods and a traditional inference method for an observation from the test set. The traditional method consists in modeling the foreground mass with a low-dimensional parametric macromodel comprising an EPL main deflector augmented with external shear and $m=3,4$ multipoles. We fix the source to the ground truth and determine the maximum likelihood macromodel parameters. Even in this idealized case, the lack of flexibility in the foreground mass model prevents modeling the observation down to the noise level, in contrast with our methods.}
    \label{fig:simulated convergence comparison traditional}
\end{figure}

Figure \ref{fig:simulated convergence reconstruction simple} shows an example of a lens model obtained for an observation from the test set. The convergence map consists of a single main deflector, making this system easier to model. We show four joint source and convergence map posterior samples. By displaying the normalized residuals between the simulated observation and the image reconstructions, as well as their p-value\footnote{defined here as the probability that a $\chi^2_k$ random variable with $k=64^2$ exceeds the $\chi^2$ value computed from the normalized residuals.}, we show that the system is modeled to the noise level.

Figure \ref{fig:simulated convergence reconstruction complex} shows another example of a lens model. The convergence map consists of two main deflectors, making this system harder to model relative to the previous one. In contrast to our methods, modeling this system with analytic profiles would entail a significant computational cost and the need for heavy user input. The normalized residuals show some structure and the corresponding p-values are reduced in comparison to the observational noise. This indicates that small biases remain in our results, particularly when modeling systems with multiple main deflectors.

In Appendix \ref{app:more results}, we show ten additional examples of lens models, with observations taken randomly from the test set. The results are consistent with the previous discussion. 

To assess the calibration of posterior samples generated with our methods, we show in Figure \ref{fig:simulated convergence tarp} a TARP (Tests of Accuracy with Random Points) plot. Calibrated TARP is a necessary and sufficient condition for a posterior estimator to be unbiased \citep{Lemos_Coogan_Hezaveh_Perreault-Levasseur_2023}. The plot is computed using 128 joint samples of the source and log convergence map for each of 128 simulated observations from the test set, and the confidence regions are computed using bootstrapping. The reference points in TARP are taken to be different source and convergence images from the test set. The test reveals only a small miscalibration, consistent with results discussed previously.

To assess the goodness of fit of the posterior samples generated with our methods, we show in Figure \ref{fig:simulated convergence chi2} the distribution of $\chi^2$ values, computed for a given simulated observation $\boldsymbol{y}$ and a posterior sample $(\boldsymbol{s},\boldsymbol{\kappa})$ as
\begin{align}
    \chi^2 = \sum_{i=1}^{64}\sum_{j=1}^{64} \Bigg( \frac{\boldsymbol{y}_{ij} - (f(\boldsymbol{s},\boldsymbol{\kappa}))_{ij}}{\sigma_y}\Bigg)^2
\end{align} 
where $i$ and $j$ denote pixel indices. We compute the $\chi^2$ value of one posterior sample per observation for all 2110 observations in the test set, and compare the resulting distribution to the theoretical distribution, which is the $\chi^2_k$ distribution with $k=64^2$ degrees of freedom. The small extended tail of the distribution is consistent with the results discussed previously. In addition, we show in Figure \ref{fig:simulated convergence chi2 percentile} the posterior sample for chosen $\chi^2$ percentile ranks, from lowest to highest. This, in particular, showcases samples from the entire range of goodness of fit on the test set.

To assess the ability of our methods to model complex foreground mass distributions, we show in Figure \ref{fig:simulated convergence collection complex} a collection of lens models for 12 observations from the test set selected on the basis that their convergence map is highly complex. For each observation, we show one joint source and convergence map sample. In several cases, more than two main deflectors are correctly identified. The last displayed model has the third lowest goodness of fit across the entirety of the test set.

\begin{figure}[t]
    \centering
    \includegraphics[width=1\linewidth]{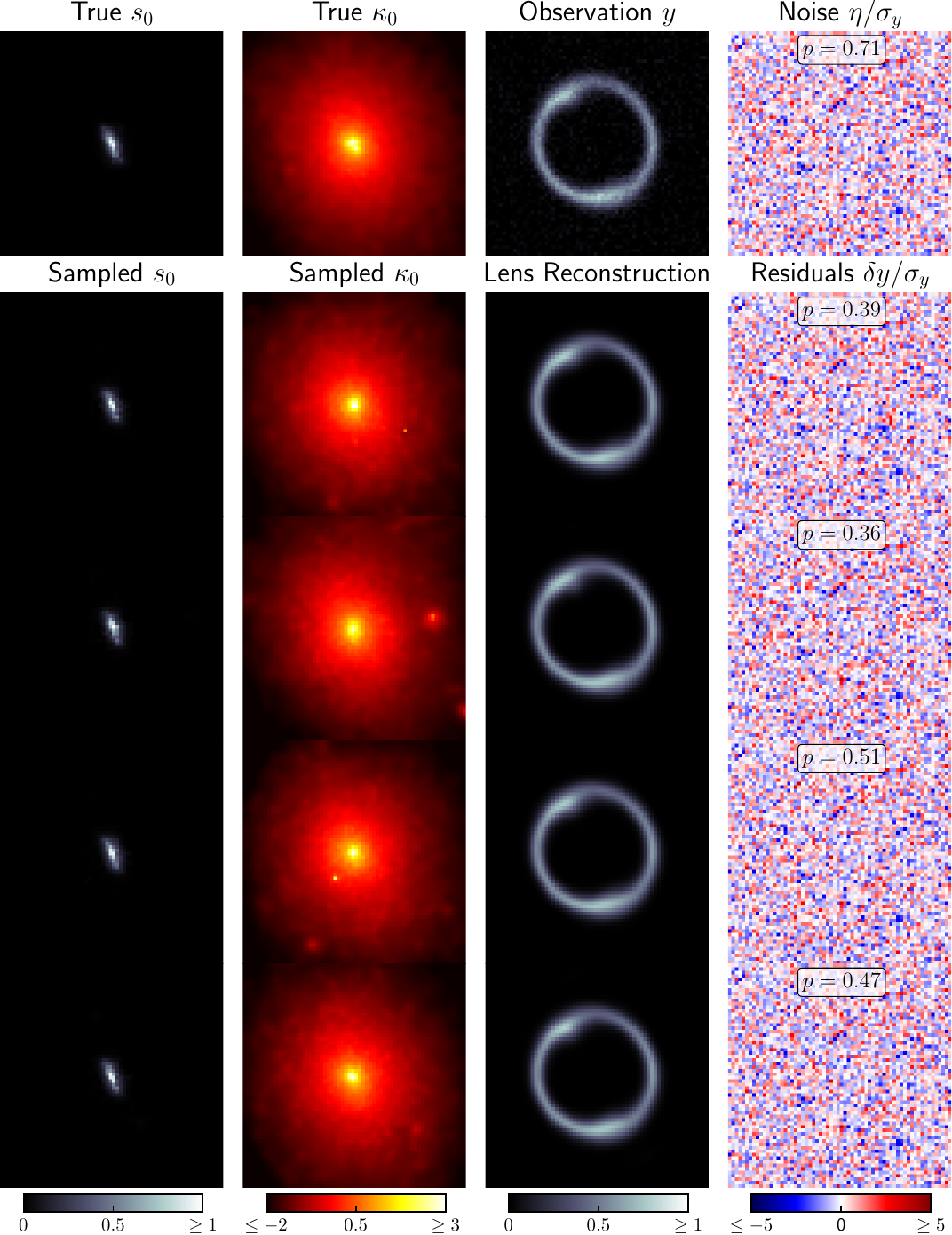}
    \caption{Test of our methods on out-of-distribution (OOD) data. The convergence map is the sum of two randomly selected convergence maps from the test set, assessing the robustness of the model to changes in the physical properties of the convergence maps. We show four joint posterior samples and the corresponding lensed image reconstruction and residuals.}
    \label{fig:simulated convergence ood}
\end{figure}

We now demonstrate the importance of the iterative refinement of posterior scores via the RIM framework in our methods. We do this by training an additional DiRIM model with the number of RIM iterations set to $M=1$, while keeping everything else fixed. This effectively disables the iterative refinement of posterior scores via the RIM framework, making it closely mimic the behavior of baseline conditional denoising score matching (see Section \ref{sec:diffusion} and specifically Figure \ref{fig:cdsm}). In Figure \ref{fig:simulated convergence comparison no rim}, we model an observation from the test set using this $M=1$ model. The resulting lensed images residuals show a lot of structure and multiple $\geq 5 \sigma_y$ pixels, indicating that the system was not modeled to the noise level. This, in particular, means that baseline conditional denoising score matching fails to accurately model strong gravitational lenses. To further demonstrate this, we add to Figure \ref{fig:simulated convergence chi2} an additional distribution of $\chi^2$ values obtained from posterior samples of the $M=1$ model. The distribution is shifted to the right and has a heavily extended tail.

In comparison to methods modeling the foreground mass distribution using parametric models, our method has additional flexibility, which translates into improved modeling of the observations. As an illustrative example, we show in Figure \ref{fig:simulated convergence comparison traditional} an example of a system for which the convergence map consists of a single main deflector, but for which a parametric fit fails to model the observation to the noise level\footnote{We find that a large fraction of systems in this dataset with a single deflector can be modeled down to the noise level using the parametric model described for the deflector, however we focus here on an example where this is not the case.}. To demonstrate this, the source image is fixed to the ground truth, and a parametric model comprising an Elliptical Power-Law (EPL) perturbed with external shear and $m=3,4$ multipoles is used to model the foreground mass. This gives a total of 12 macromodel parameters, which are described in Appendix \ref{app:macromodel}. The parameters of the macromodel are determined by maximizing the likelihood of the reconstructed image given the data. The residuals between the observation and image reconstruction indicate a bias caused by the lack of flexibility of the parametric model. This is in contrast to our method, which models the observation down to the noise level.

\begin{figure}[t]
    \centering
    \includegraphics[width=1\linewidth]{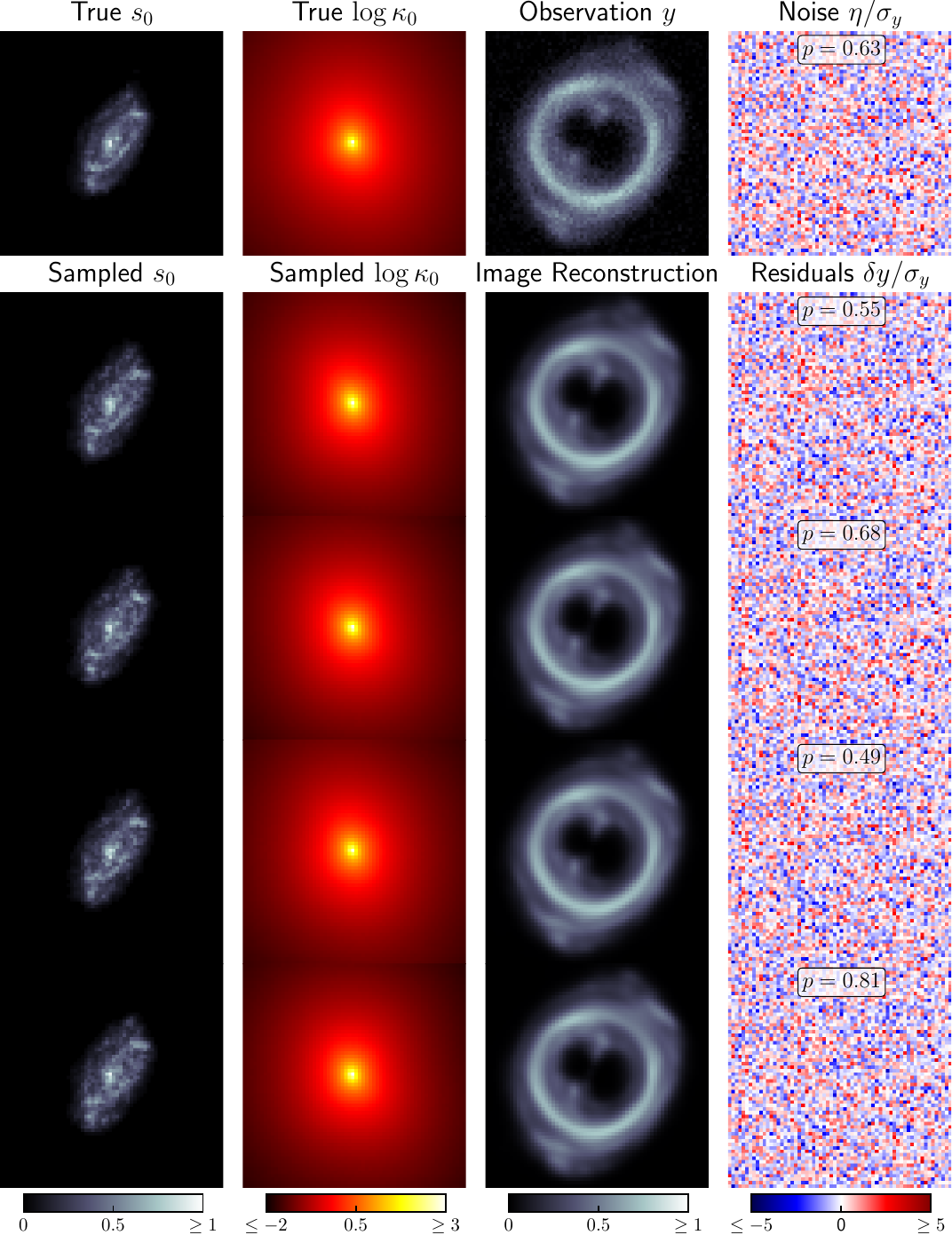}
    \caption{Lens model for a mock observation from the test set. The convergence map is generated from the analytic profile, and does not contain a subhalo. We show the true source image, the true convergence map, the mock observation, and four joint source and convergence map samples, along with the reconstructed lensed image and residuals for each.}
    \label{fig:analytic convergence reconstruction simple}
\end{figure}

\begin{figure}[t]
    \centering
    \includegraphics[width=1\linewidth]{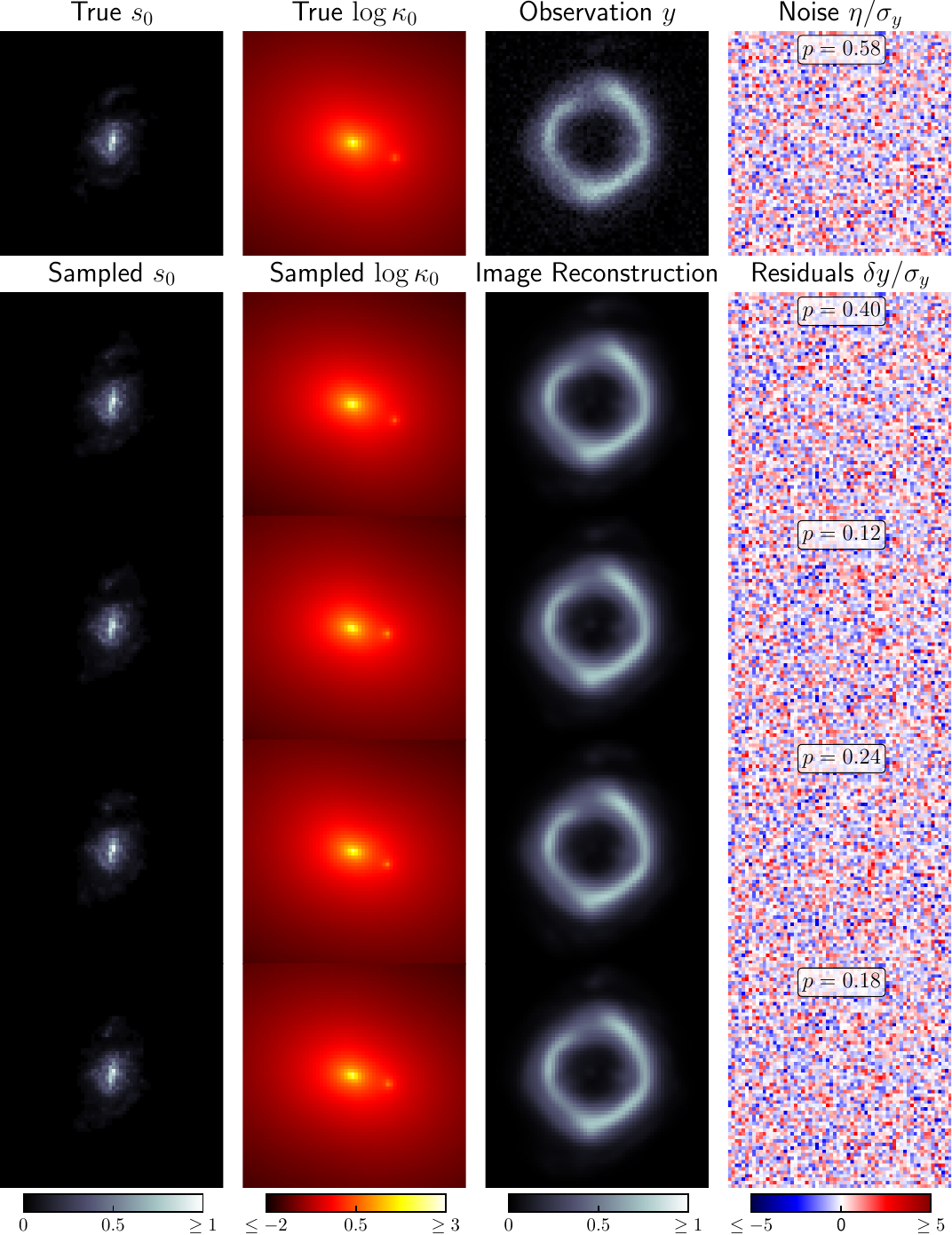}
    \caption{Same as Figure \ref{fig:analytic convergence reconstruction simple} for a different mock observation for which the convergence map contains a subhalo.}
    \label{fig:analytic convergence reconstruction complex}
\end{figure}

\begin{figure*}[t]
    \centering
    \includegraphics[width=1\linewidth]{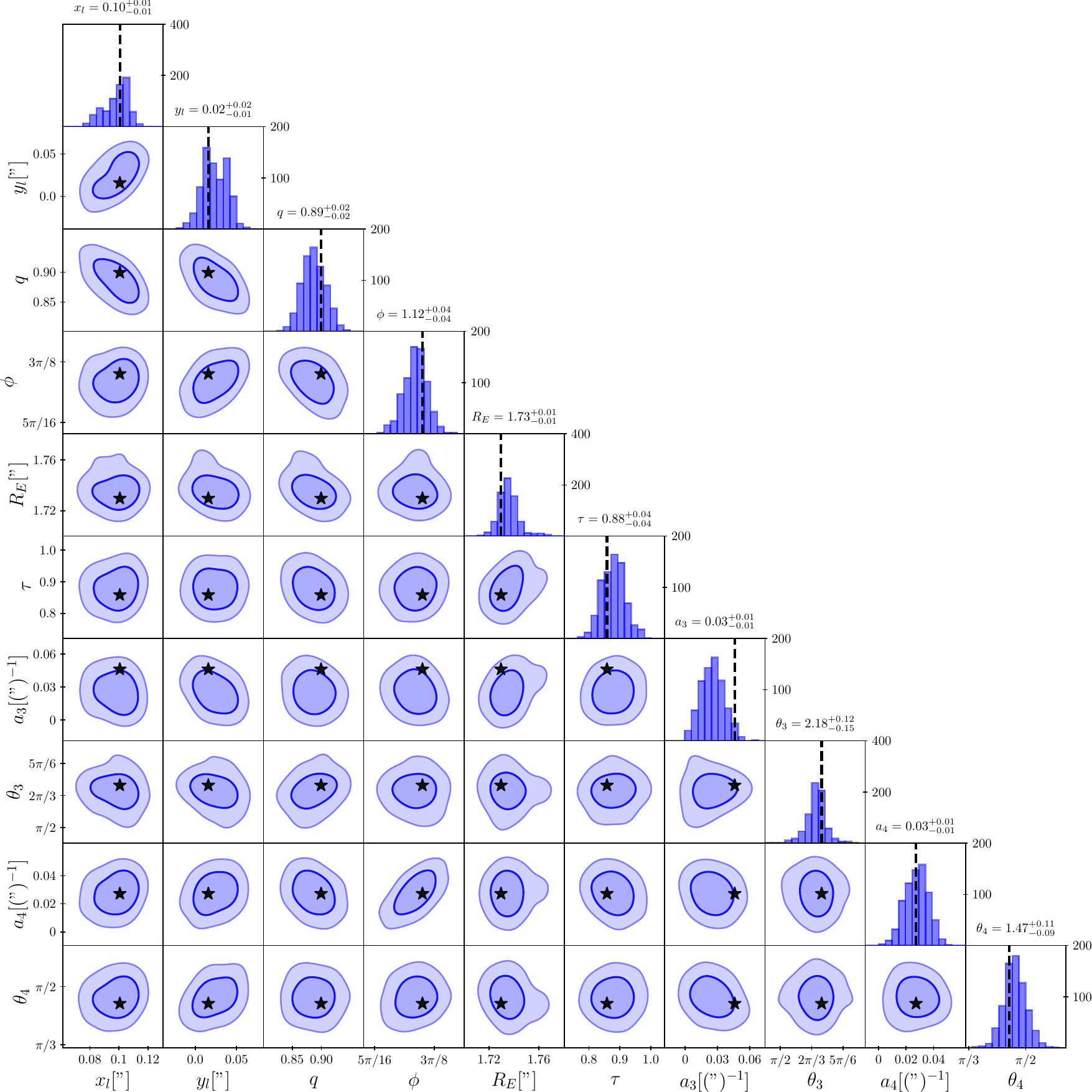}
    \caption{Posterior distribution of the macromodel parameters used to generate the analytic convergence maps for a mock observation from the test set. The posterior samples are obtained by performing a least-squares fit on the pixelated convergence map samples obtained with our methods. The true macromodel parameters used to generate the convergence map are shown as black stars and black dotted lines. The dark and light shaded region correspond to the $1\sigma$ and $2\sigma$ confidence regions, respectively.}
    \label{fig:analytic convergence fit macro parameters}
\end{figure*}

Finally, to assess the ability of DiRIM to generalize to out-of-distribution (OOD) data, we show in Fig.\ref{fig:simulated convergence ood} an example of a lens model for which the convergence map is out-of-distribution with respect to the training set. We take the convergence map to be the sum of two randomly selected convergence maps from the test set. This results in a convergence map with two co-located halos that have not interacted with each other. The ability of DiRIM to model this system shows robustness to changes in the physical properties of the convergence maps.

 \subsection{Analytic convergence model}

This section discusses our results on tests done with the DiRIM model trained on the set of analytic convergence maps. In comparison to the previous results subsection, the analytic nature of the convergence maps in this subsection allows for several additional consistency tests of our methods.

Figure \ref{fig:analytic convergence reconstruction simple} shows an example of a lens model obtained for an observation from the test set. In this example, the convergence map does not contain a subhalo. By displaying the normalized residuals between the simulated observation and the image reconstructions, as well as their p-value, we show that the system is modeled to the noise level.

Figure \ref{fig:analytic convergence reconstruction complex} shows another example of a lens model obtained for an observation from the test set. In this example, the convergence map contains a subhalo. The four displayed sampled convergence maps also contain the subhalo, signifying its detection. While Figure \ref{fig:analytic convergence reconstruction complex} shows only 4 posterior samples, we have generated 740 posterior samples and verified that all 740 convergence maps contain the subhalo. This corresponds to a $\gsim 3\sigma$ detection of the subhalo. The normalized residuals show some structure and the corresponding p-values are reduced in comparison to the noise used to generate the simulated observation, indicating that the observation is modeled with a small bias. To demonstrate the manner in which the RIM in our framework enables the detection of this subhalo, we show in App.\ref{app:demo rim iterations} examples of the evolution of the source and convergence map images through RIM iterations during the denoising processes that were done to generate the samples in Figure \ref{fig:analytic convergence reconstruction complex}.

\begin{figure}[t]
    \centering
    \includegraphics[width=1\linewidth]{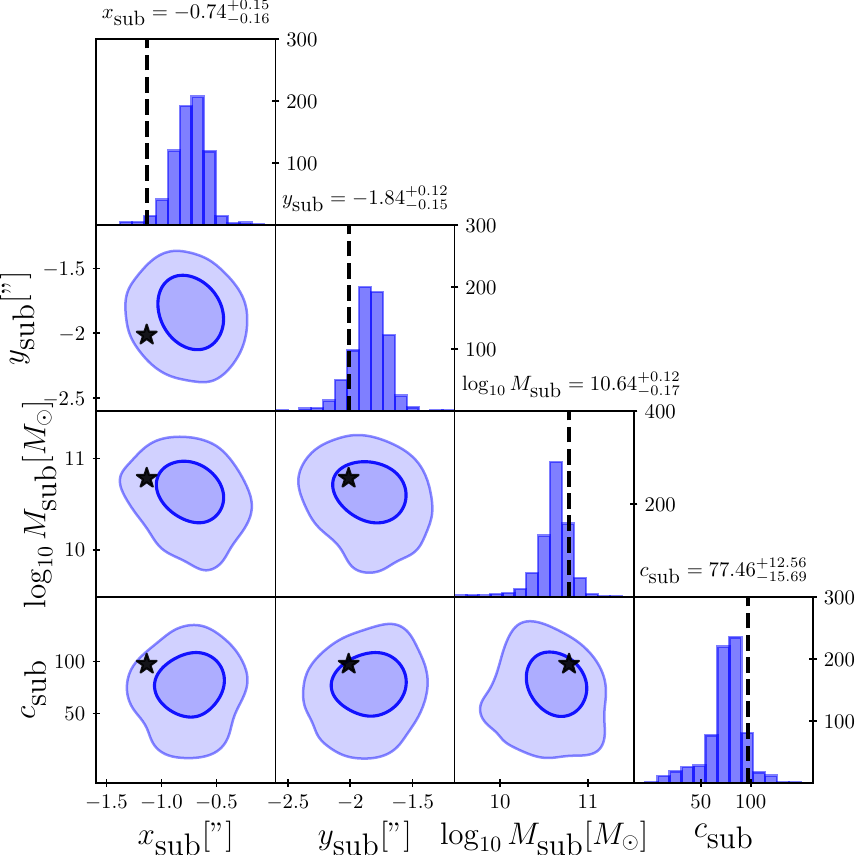}
    \caption{Same as Figure \ref{fig:analytic convergence fit macro parameters} except we show the subhalo parameters as opposed to the main deflector parameters.}
    \label{fig:analytic convergence fit subhalo parameters}
\end{figure}

While in our methods foreground mass distributions are represented as pixelated images, we show that we can also recover, from simulated observations, the macromodel parameters used to generate the underlying convergence map. To do so, we generate joint posterior samples of the source and convergence map and fit, for each sample, all macromodel and subhalo parameters in Table \ref{tab:priors} to the sampled pixelated convergence map via a least-squares fit. In Figure \ref{fig:analytic convergence fit macro parameters} we show, for an observation from the test set, the resulting posterior distribution on the macromodel parameters, while in Figure \ref{fig:analytic convergence fit subhalo parameters} we show the posterior distribution on the subhalo parameters. The figures are generated using 740 posterior samples\footnote{A few posterior samples do not contain the subhalo, and these samples are excluded from the fit. Additionally, for visualization purposes, the angular coordinates $\theta_3$, $\theta_4$ are rewrapped to avoid splitting their posterior at their periodic boundary.}. The true macromodel and subhalo parameters used to generate the convergence map are also displayed and are consistent with the posteriors. We repeat this test for 20 observations from the test set for which a subhalo is present in the convergence map and detected in at least 98\% of 740 samples, and we report in Table \ref{tab:subhalo parameters} the mean uncertainty on the subhalo parameters.

\begin{figure}[t]
    \centering
    \includegraphics[width=1\linewidth]{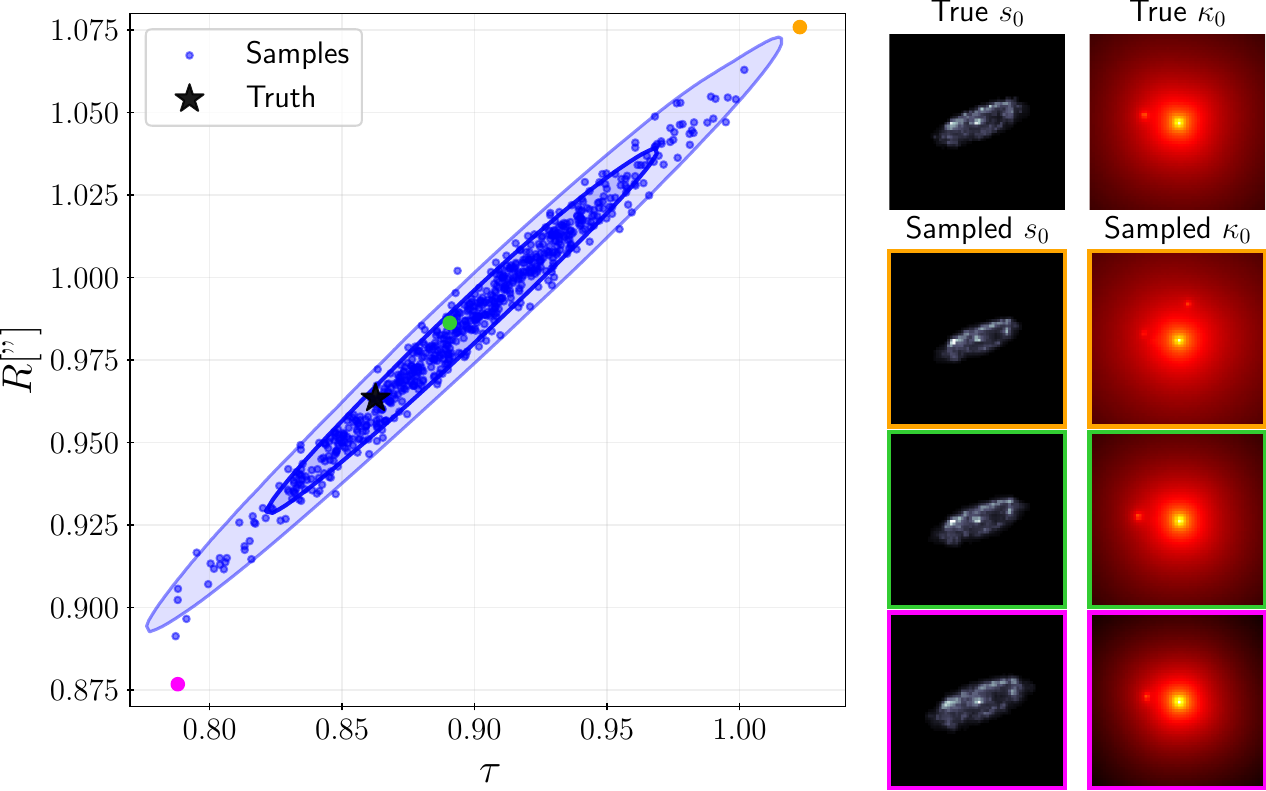}
    \caption{Coverage of the degeneracy between the size of the source $R$ and the radial slope of the convergence map $\tau$ across posterior samples. The blue points and confidence contours are computed from posterior samples for an observation from the test set. The orange, green and pink points correspond to the samples with the smallest, median, and largest value of $R$, respectively, and their associated source and convergence map joint samples are shown on the right. The value of $R$ and $\tau$ corresponding to the true source and convergence map is shown as a black star.}
    \label{fig:analytic convergence source convergence degeneracy}
\end{figure}

\begin{table}
    \centering
    \begin{tabular}{c||c}
        \hline \hline
        Parameter & Average $\sigma$\\
        \hline
        $x_{\text{sub}}['']$ & 0.136\\
        $y_{\text{sub}}['']$ & 0.145\\
        $\log_{10} M_{\text{sub}}[M_\odot]$ & 0.167\\
        $c_{\text{sub}}$ & 16.7\\
    \end{tabular}
    \caption{Mean uncertainty on fitted subhalo parameters across 20 observations from the test set. The uncertainty on each parameter is calculated as the standard deviation of the corresponding marginal posterior distribution.}
    \label{tab:subhalo parameters}
\end{table}

\begin{figure}[t]
    \centering
    \includegraphics[width=1\linewidth]{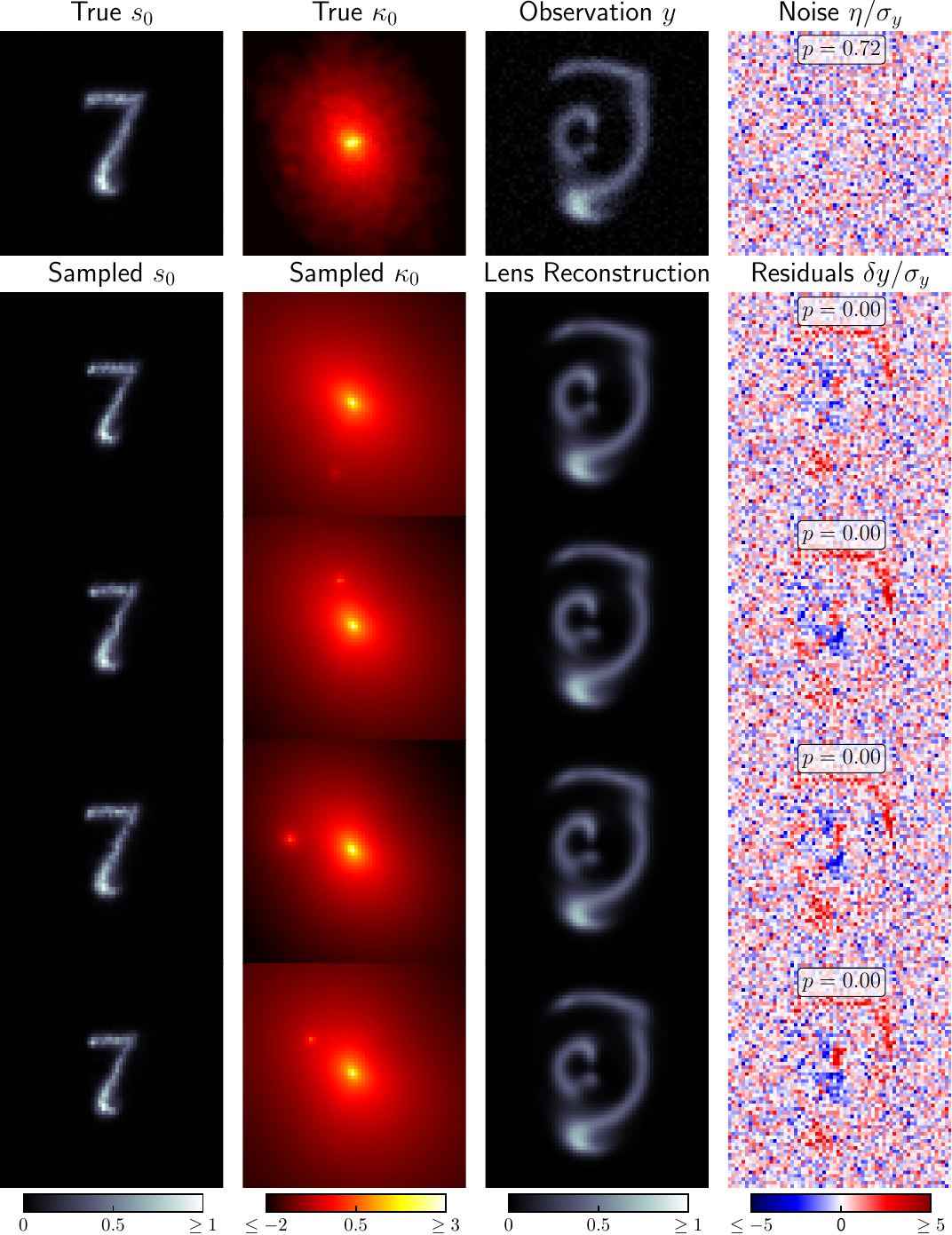}
    \caption{Test of our methods on out-of-distribution (OOD) data. The source image is an AI-generated image of a galaxy in the shape of the number $7$. The convergence map is taken from the set of simulated convergence maps, while the model is trained on the set of analytic convergence maps. We show four joint posterior samples and the corresponding lensed image reconstruction and residuals.}
    \label{fig:analytic convergence ood}
\end{figure}

In figures of this paper showing joint source and convergence map samples (see e.g., Figure \ref{fig:simulated convergence reconstruction simple}), the difference in the size of the source between individual samples is a demonstration of the ability of DiRIM to sample posteriors with degeneracies, in this case, the degeneracy between the size of the source and the radial slope of the convergence map. To better demonstrate this, we show in Figure \ref{fig:analytic convergence source convergence degeneracy} a collection of samples from the marginal distribution $p(R,\tau|\boldsymbol{y})$ where $R$ is the size of the source and $\tau$ is the slope of the convergence map. The posterior appears as an elongated ellipse, demonstrating that the degenerate direction in $(R,\tau)$ space is appropriately sampled. We also show the joint source and convergence map posterior samples with the smallest, median, and largest values of $R$, allowing to see by eye the variation of the size of the source and slope of the convergence map. The size of the source $R$ in arcseconds is computed by considering the brightness distribution of a source image as a normalized probability distribution and computing its absolute first moment\footnote{When computing $R$ we clip all source pixels below $0.03$ to $0$ to avoid $R$ being dominated by faint features at the extremities of the source images. This is also why we define $R$ in terms of the absolute first moment rather than the second moment.}.

To further assess the capacity of DiRIM to generalize to OOD data, we show in Figure \ref{fig:analytic convergence ood} an example of a lens model for which the source image and convergence map image are both out-of-distribution with respect to the training set. Inspired by \cite{Adam_Coogan_Malkin_Legin_Perreault-Levasseur_Hezaveh_Bengio_2022}, we take as the source image an AI-generated image of a galaxy in the shape of the number $7$, while the convergence map image is taken from the set of simulated convergence maps described in Section \ref{subsubsec:convergence data}. We show four joint source and convergence map posterior samples. While the reconstructed image residuals show significant structure and have heavily reduced p-values, the morphology of the source and convergence map is inferred correctly, demonstrating some capacity to generalize to OOD data.

\begin{figure}[t]
    \centering
    \includegraphics[width=1\linewidth]{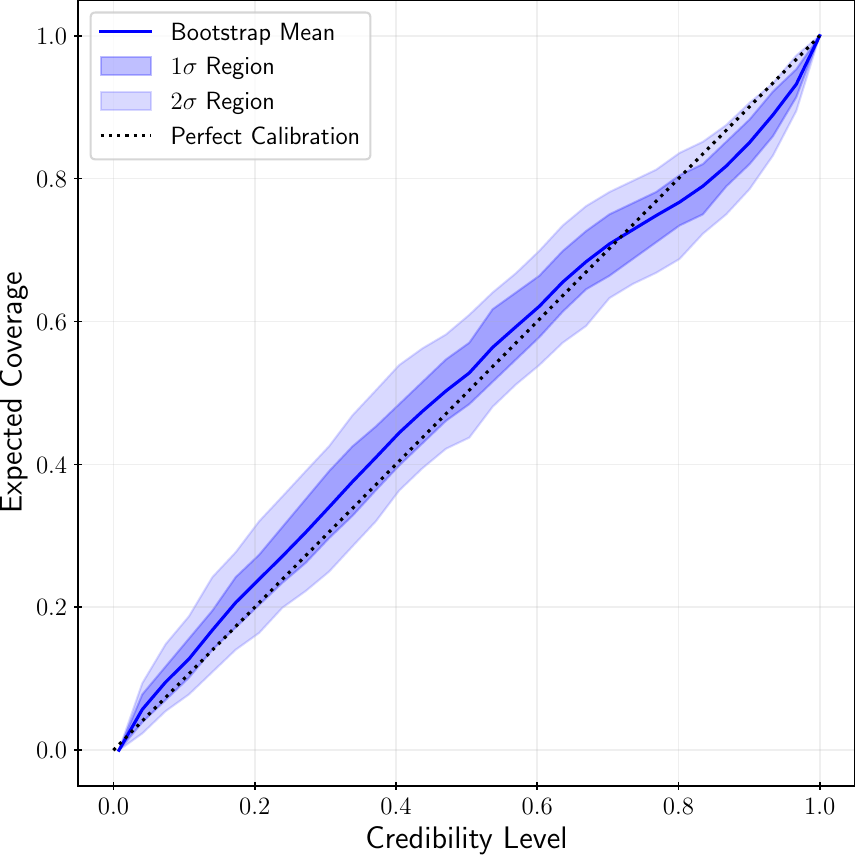}
    \caption{Tests of Accuracy with Random Points (TARP) plot demonstrating the near-perfect calibration of our posterior samples in pixel-space on the test set.}
    \label{fig:analytic convergence tarp}
\end{figure}

To assess the calibration of posterior samples generated with our methods, we show in Figure \ref{fig:analytic convergence tarp} a TARP plot generated in the same way as the previously discussed TARP plot (Figure \ref{fig:simulated convergence tarp}). The test reveals only a small miscalibration.

\section{Conclusion} \label{sec:conclusion}

We have presented DiRIM, a first-of-its-kind method for pixel-space joint posterior sampling of the source and foreground mass distribution in strong gravitational lensing. Overall, our results demonstrate that DiRIM can leverage highly flexible representations of the source and foreground mass to model, in a Bayesian framework, realistic gravitational lensing observations down to the noise level.

In the future, we plan to address the small remaining biases present in our results. We also plan to apply this framework to other high-dimensional non-linear inverse problems. It is potentially applicable to any inverse problem for which the forward model can be made differentiable and computationally inexpensive. Additionally, motivated by the success of latent diffusion models, we plan to investigate a latent-space formulation of DiRIM, which would allow higher resolution generation at a reduced computational cost.

\section*{Software and Data} The source code and scripts used in this work are publicly available as a Python package under the name DiRIM\_Lensing\footnote{\href{https://github.com/GuillaumePayeur/DiRIM_Lensing}{https://github.com/GuillaumePayeur/DiRIM\_Lensing}}. The source and convergence map images used during training and testing are available and hosted on Zenodo\footnote{\href{https://zenodo.org/records/19668019}{https://zenodo.org/records/19668019}}.

This work made use of the packages {\textsc{Caustics} \citep{Stone_Adam_Coogan_Yantovski-Barth_Filipp_Setiawan_Core_Legin_Wilson_Barco_et_al._2024}, \textsc{Pytorch} \citep{Paszke_Gross_Massa_Lerer_Bradbury_Chanan_Killeen_Lin_Gimelshein_Antiga_et_al._2019}, \textsc{Numpy} \citep{Harris_Millman_Walt_Gommers_Virtanen_Cournapeau_Wieser_Taylor_Berg_Smith_et_al._2020}, \textsc{TARP} \citep{Lemos_Coogan_Hezaveh_Perreault-Levasseur_2023}, \textsc{Matplotlib} \citep{Hunter_2007} and \textsc{Pydantic} \citep{Colvin_Pydantic_Validation_2026}.}

\section*{Acknowledgments}

The authors thank Alexandre Adam, Nicolas Payot, Ronan Legin and Huilin Tai for helpful discussions and comments. This work is supported in part by computational resources provided by Calcul Quebec and the Digital Research Alliance of Canada.  This work is partially supported by Schmidt Futures, a philanthropic initiative founded by Eric and Wendy Schmidt as part of the Virtual Institute for Astrophysics (VIA). G.P. acknowledges the support of the Natural Sciences and Engineering Research Council of Canada (NSERC) [funding reference number 601521]. G.M.B. acknowledges support from the Fonds de recherche du Québec – Nature et technologies (FRQNT) under a Doctoral Research Scholarship (\doi{10.69777/368273}). Y.H. and L.P. acknowledge support from the Canada Research Chairs Program, the National Sciences and Engineering Council of Canada through grants RGPIN-2020-05073 and 05102.

\bibliography{references}{}
\bibliographystyle{aasjournalv7}

\appendix

\section{derivations concerning diffusion} \label{app:proofs}

\subsection{Relation between score matching and denoising} \label{app:sm denoising equivalence}

For the VE SDE Eq.\ref{eq:VE SDE}, one can show \citep{Lai_Song_Kim_Mitsufuji_Ermon_2025} that the perturbation kernel $p(\boldsymbol{x}_t|\boldsymbol{x}_0)$ is given by
\begin{align}
    p_t(\boldsymbol{x}_t|\boldsymbol{x}_0) = \mathcal{N}(\boldsymbol{x}_t;\boldsymbol{x}_0, \sigma^2(t)I),
\end{align}
and thus
\begin{align}
    \nabla_{\boldsymbol{x}_t}\log p_t(\boldsymbol{x}_t|\boldsymbol{x}_0) = \frac{\boldsymbol{x}_0-\boldsymbol{x}_t}{\sigma^2(t)}.
\end{align}
Denoting
\begin{align}
    \mathbb{E} \equiv \mathbb{E}_{t\sim \mathcal{U}[0,T],\boldsymbol{x}_0\sim p(\boldsymbol{x}),\boldsymbol{y}\sim p(\boldsymbol{y}|\boldsymbol{x}_0),\boldsymbol{x}_t\sim p(\boldsymbol{x}_t|\boldsymbol{x}_0)}, \label{eq:shorthand E}
\end{align}
we can thus write the loss $L_{\text{CDSM}}$ in Eq.\ref{eq:loss CDSM explicit} as
\begin{align}
    L_{\text{CDSM}} &= \mathbb{E}\bigg\{\lambda(t) \Big\|\boldsymbol{s}_\theta(\boldsymbol{x}_t,\boldsymbol{y},t)-\frac{\boldsymbol{x}_0-\boldsymbol{x}_t}{\sigma^2(t)}\Big\|_2^2\bigg\} \nonumber \\
    &= \mathbb{E}\bigg\{\frac{\lambda(t)}{\sigma^4(t)} \|\boldsymbol{s}_\theta(\boldsymbol{x}_t,\boldsymbol{y},t)\sigma^2(t)+\boldsymbol{x}_t-\boldsymbol{x}_0\|_2^2\bigg\} \nonumber \\
    &= \mathbb{E}\bigg\{\frac{\lambda(t)}{\sigma^4(t)} \|\boldsymbol{g}_\theta(\boldsymbol{x}_t,\boldsymbol{y},t)-\boldsymbol{x}_0\|_2^2\bigg\},
\end{align}
which is Eq.\ref{eq:loss CDSM}.

\subsection{Relation between DiRIM loss and CDSM loss} \label{app:sm DiRIM and CDSM loss equivalence}

The minimizer of the DiRIM loss Eq.\ref{eq:loss DiRIM}, using again the notation of Eq.\ref{eq:shorthand E}, is
\begin{align}
    \argmin_\theta L_{\text{DiRIM}} = \argmin_\theta \mathbb{E}\bigg\{ W(t) \sum_{m=1}^{M} w_m \|\hat{\boldsymbol{x}}_0^{(m)}-\boldsymbol{x}_0\|_2^2\bigg\}. 
\end{align}
For the choice of weights $w_m$ in Eq.\ref{eq:weights w_m}, this reads
\begin{align}
    \argmin_\theta L_{\text{DiRIM}} &= \argmin_\theta \mathbb{E}\{W(t)\|\hat{\boldsymbol{x}}_0^{(M)}-\boldsymbol{x}_0\|_2^2\} \nonumber \\
    &= \argmin_\theta \mathbb{E}\bigg\{\frac{\lambda(t)}{\sigma^4(t)}\|\hat{\boldsymbol{x}}_0^{(M)}-\boldsymbol{x}_0\|_2^2\bigg\} \nonumber \\
    &= \argmin_\theta L_{\text{CDSM}},
\end{align}
where the last line follows from identifying $\boldsymbol{g}_\theta(\boldsymbol{x}_t,\boldsymbol{y},t)\equiv \hat{\boldsymbol{x}}_0^{(M)}$.

\subsection{DiRIM with the VP SDE} \label{app:DiRIM VP}

For the variance preserving (VP) SDE \citep{Song_Sohl-Dickstein_Kingma_Kumar_Ermon_Poole_2020} defined by
\begin{align}
    d\boldsymbol{x} = -\frac{1}{2}\beta(t)\boldsymbol{x} dt+\sqrt{\beta(t)}d\boldsymbol{w}, \label{eq:VP SDE}
\end{align}
the equivalent of Eq.\ref{eq:posterior score DiRIM VE SDE} is
\begin{align}
    \nabla_{\boldsymbol{x}_t}\log p_t(\boldsymbol{x}_t|\boldsymbol{y}) = \frac{\boldsymbol{x}_0^{(M)}\alpha(t)-\boldsymbol{x}_t}{\sigma^2(t)}, \label{eq:posterior score DiRIM VP SDE}
\end{align}
where $\alpha(t)$ and $\sigma(t)$ are defined as
\begin{align}
    \alpha(t) &= \exp \Big(-\frac{1}{2}\int_0^t\beta(s)ds\Big) \nonumber \\
    \sigma(t) &= \sqrt{1-\alpha^2(t)}. \label{eq:alpha sigma}
\end{align}
To see this, consider that for the VP SDE Eq.\ref{eq:VP SDE}, the perturbation kernel $p_t(\boldsymbol{x}_t|\boldsymbol{x}_0)$ is given by \citep{Lai_Song_Kim_Mitsufuji_Ermon_2025}
\begin{align}
    p_t(\boldsymbol{x}_t|\boldsymbol{x}_0) = \mathcal{N}(\boldsymbol{x}_t|\boldsymbol{x}_0\alpha(t),\sigma^2(t) I),
\end{align}
with $\alpha(t)$ and $\sigma(t)$ defined in Eq.\ref{eq:alpha sigma}. Thus,
\begin{align}
    \nabla_{\boldsymbol{x}_t}\log p_t(\boldsymbol{x}_t|\boldsymbol{x}_0) = \frac{\boldsymbol{x}_0\alpha(t)-\boldsymbol{x}_t}{\sigma^2(t)}.
\end{align}
Using again the notation of Eq.\ref{eq:shorthand E}, we can thus write the loss $L_{\text{CDSM}}$ in Eq.\ref{eq:loss CDSM explicit} as
\begin{align}
    L_{\text{CDSM}} &= \mathbb{E}\bigg\{\lambda(t) \Big\|\boldsymbol{s}_\theta(\boldsymbol{x}_t,\boldsymbol{y},t)-\frac{\boldsymbol{x}_0\alpha(t)-\boldsymbol{x}_t}{\sigma^2(t)}\Big\|_2^2\bigg\} \nonumber \\
    &= \mathbb{E}\bigg\{\frac{\lambda(t)\alpha^2(t)}{\sigma^4(t)} \Big\|\frac{\boldsymbol{s}_\theta(\boldsymbol{x}_t,\boldsymbol{y},t)\sigma^2(t)+\boldsymbol{x}_t}{\alpha(t)}-\boldsymbol{x}_0\Big\|_2^2\bigg\} \nonumber \\
    &= \mathbb{E}\bigg\{\frac{\lambda(t)\alpha^2(t)}{\sigma^4(t)} \Big\|\boldsymbol{g}_\theta(\boldsymbol{x}_t,\boldsymbol{y},t)-\boldsymbol{x}_0\Big\|_2^2\bigg\},
\end{align}
where $\boldsymbol{g}_\theta$ and $\boldsymbol{s}_\theta$ are related by
\begin{align}
    \boldsymbol{s}_\theta = \frac{\boldsymbol{g}_\theta \alpha(t)-\boldsymbol{x}_t}{\sigma^2(t)}.
\end{align}
As in the VE case, identifying $\boldsymbol{g}_\theta(\boldsymbol{x}_t,\boldsymbol{y},t) \equiv \hat{\boldsymbol{x}}_0^{(M)}$, we conclude that we can obtain posterior scores via
\begin{align}
    \nabla_{\boldsymbol{x}_t}\log p_t(\boldsymbol{x}_t|\boldsymbol{y}) = \frac{\boldsymbol{x}_0^{(M)}\alpha(t)-\boldsymbol{x}_t}{\sigma^2(t)},
\end{align}
which is Eq.\ref{eq:posterior score DiRIM VP SDE}.

\section{Choice of diffusion time weights in the loss function} \label{app:lambda}

In this work, the function $W(t)$ in the loss function Eq.\ref{eq:loss DiRIM} is chosen by considering two factors:
\begin{enumerate}
    \item For $t \approx 1$, the denoising neural network $g_\theta(\boldsymbol{x}_t,\boldsymbol{y},t)$ is mostly informed by the observation $\boldsymbol{y}$ over the noisy source and convergence map $\boldsymbol{x_t}$, since the latter is almost pure noise. Hence, the function $W(t)$ is chosen to be constant near $t = 1$.
    \item For $t \approx 0$, the denoising neural network $g_\theta(\boldsymbol{x}_t,\boldsymbol{y},t)$ is mostly informed by the noisy source and convergence map $\boldsymbol{x}_t$ over the observation $\boldsymbol{y}$, since the former is almost noiseless. Hence, the function $W(t)$ is chosen to satisfy $W(t) \propto \alpha(t)^2/\sigma(t)^2$, which is standard for unconditional diffusion models (see e.g., \citealt{Song_Sohl-Dickstein_Kingma_Kumar_Ermon_Poole_2020}).
\end{enumerate}
$W(t)$ is therefore set to
\begin{align}
    W(t) = \frac{\bar{\sigma}^2}{\min\{\bar{\sigma}^2,\sigma(t)^2/\alpha^2(t)\}},
\end{align}
where $\bar{\sigma}$ is a free parameter which governs at which point in time $W(t)$ transitions between its large $t$ constant phase and its low $t$ phase. We set $\bar{\sigma} = 2\times 10^{-2}$ for the model trained on analytic convergence maps and $\bar{\sigma} = 4\times 10^{-2}$ for the model trained on simulated convergence maps since we found that these values perform best empirically.

\section{SDE implementation details} \label{app:solving SDE}

In this work, we use the VP SDE Eq.\ref{eq:VP SDE} with the noise schedule $\beta(t)$ given by
\begin{align}
    \beta(t) = \beta_{\text{min}}\Big(\frac{\beta_{\text{max}}}{\beta_{\text{min}}}\Big)^t,
\end{align}
with $\beta_\text{min} = 2 \times 10^{-3}$ and $\beta_{\text{max}} = 250$. To avoid numerical instabilities at $t=0$ during training and sampling \citep{Song_Sohl-Dickstein_Kingma_Kumar_Ermon_Poole_2020}, we train the DiRIM only on the time interval $[\epsilon,1]$ with $\epsilon=10^{-3}$.

To solve the reverse SDE from $t=1$ to $t=\epsilon$, we use the Euler-Maruyama (EM) method \citep{maruyama1955continuous, kloeden1992numerical} with 1000 steps. After these steps, we pass the sampled source and convergence map through the denoising model to obtain the finalized samples. This effectively projects the samples from $t=\epsilon$ to $t=0$ in one step.

\section{U-Net architecture and training procedure} \label{app:unet}

In this work, the architecture of the U-Net $\boldsymbol{g}_\theta$ is the one discussed in \cite{Karras_Aittala_Aila_Laine_2022}. We adapt its implementation found in the \textsc{edm}\footnote{https://github.com/nvlabs/edm} Python package. Our choice of hyperparameters when training the model is detailed in Table \ref{tab:hyperparameters unet}. We use slightly different hyperparameters for the model trained on analytic convergence maps and for the model trained on simulated convergence maps. Both sets of parameters are reported in the table.

\begin{table*}
    \centering
    \caption{U-Net hyperparameters used in this work.}
    \begin{tabular}{c||c||c}
        \hline \hline
        Hyperparameter & Value (simulated convergence maps model) & Value (analytic convergence maps model) \\
        \hline
        Levels in encoder \& decoder & 5 & 5\\
        Channels per level & \{64, 128, 256, 512, 512\} & \{128, 256, 512, 512, 512\}\\
        Residual blocks per level & 2 & 2\\
        Resolutions with attention & 4, 8 & None\\
        Number of trainable parameters & 112,264,706 & 154,317,314\\
    \end{tabular}
    \label{tab:hyperparameters unet}
\end{table*}

During training and at inference time, the likelihood gradients $\nabla_{\hat{\boldsymbol{s}}_0^{(m)}}\mathcal{L}$ and $\nabla_{\hat{\boldsymbol{\kappa}}_0^{(m)}}\mathcal{L}$ ($\nabla_s \mathcal{L}$ and $\nabla_\kappa \mathcal{L}$ for short) are pre-processed with a $\tanh$ function as
\begin{align}
    \nabla_s \mathcal{L} \rightarrow \tanh(\nabla_s \mathcal{L}/c_s)\\
    \nabla_\kappa \mathcal{L} \rightarrow \tanh(\nabla_\kappa \mathcal{L}/c_\kappa),
\end{align}
where $c_s$ and $c_\kappa$ are normalization factors treated as hyperparameters. Their value and other RIM hyperparameters are detailed in Table \ref{tab:hyperparameters rim}.

\begin{figure*}
    \centering
    \definecolor{mygreen}{HTML}{267326}
  
  \begin{tikzpicture}[node distance=2cm]
    \node (x^0) {$(\hat{\boldsymbol{s}}_0^{(0)},\log \hat{\boldsymbol{\kappa}}_0^{(0)})$};
    % \node (h^0)  [below of=x^0, node distance=0.75cm, xshift=0cm] {$\boldsymbol{h}^{(0)}$};
    \node (y^0) [above of=x^0, node distance =1cm] {$\hat{\boldsymbol{y}}^{(0)}$};
    \node (nabla_x^0) [right of=y^0, node distance =1.75cm, yshift=0.25cm] {$\nabla_{(\hat{\boldsymbol{s}}_0^{(0)},\hat{\boldsymbol{\kappa}}_0^{(0)})} \mathcal{L}$};    
    \node (res^0) [right of=y^0, node distance =1.5cm, yshift=-0.25cm] {$\delta\boldsymbol{y}^{(0)}$};  
    \node (g) [draw, rounded corners, right of=x^0, node distance=2cm] {$\boldsymbol{g}_{\theta}$};
    \node (xt0) [below of=g, node distance=0.75cm, xshift=-0.9cm] {$(\boldsymbol{s}_t,\log\boldsymbol{\kappa}_t)$};
    \node (y0) [below of=g, node distance=0.75cm, xshift=0.3cm] {$\boldsymbol{y}$};
    \node (t0) [below of=g, node distance=1.1cm] {$\log t$};
    \node(+0) [draw, rounded corners, right of=g, node distance=0.75cm] {$+$};

    \node (x^1) [right of= +0, node distance=1.75cm] {$(\hat{\boldsymbol{s}}_0^{(1)},\log \hat{\boldsymbol{\kappa}}_0^{(1)})$};
    % \node (h^1)  [below of=x^1, node distance=0.75cm, xshift=0cm] {$\boldsymbol{h}^{(1)}$};
    \node (y^1) [above of=x^1, node distance =1cm] {$\hat{\boldsymbol{y}}^{(1)}$};
    \node (nabla_x^1) [right of=y^1, node distance =1.75cm, yshift=0.25cm] {$\nabla_{(\hat{\boldsymbol{s}}_0^{(1)},\hat{\boldsymbol{\kappa}}_0^{(1)})} \mathcal{L}$};    
    \node (res^1) [right of=y^1, node distance =1.5cm, yshift=-0.25cm] {$\delta\boldsymbol{y}^{(1)}$};  
    \node (g1) [right of=x^1, node distance=2cm] {$\cdots$};
    \node (xt1) [below of=g1, node distance=0.75cm, xshift=-0.9cm] {$(\boldsymbol{s}_t,\log\boldsymbol{\kappa}_t)$};
    \node (y1) [below of=g1, node distance=0.75cm, xshift=0.3cm] {$\boldsymbol{y}$};
    \node (t1) [below of=g1, node distance=1.1cm] {$\log t$};

    \node(L^1) [below of=x^1, node distance=1cm] {$L^{(1)}$};

    \node (x^M-1) [right of=g1, node distance=2.5cm] {$(\hat{\boldsymbol{s}}_0^{(M-1)},\log \hat{\boldsymbol{\kappa}}_0^{(M-1)})$};
    % \node (h^M-1)  [below of=x^M-1, node distance=0.75cm] {$\boldsymbol{h}^{(M-1)}$};
    \node (y^M-1) [above of=x^M-1, node distance =1cm] {$\hat{\boldsymbol{y}}^{(M-1)}$};
    \node (nabla_x^M-1) [right of=y^M-1, node distance =2.5cm, yshift=0.25cm] {$\nabla_{(\hat{\boldsymbol{s}}_0^{(M-1)},\hat{\boldsymbol{\kappa}}_0^{(M-1)})} \mathcal{L}$};    
    \node (res^M-1) [right of=y^M-1, node distance =2cm, yshift=-0.25cm] {$\delta\boldsymbol{y}^{(M-1)}$};  
    \node (g^M-1) [draw, rounded corners, right of=x^M-1, node distance=2.5cm] {$\boldsymbol{g}_{\theta}$};
    \node (xtM-1) [below of=g^M-1, node distance=0.75cm, xshift=-0.9cm] {$(\boldsymbol{s}_t,\log\boldsymbol{\kappa}_t)$};
    \node (yM-1) [below of=g^M-1, node distance=0.75cm, xshift=0.3cm] {$\boldsymbol{y}$};
    \node (tM-1) [below of=g^M-1, node distance=1.1cm] {$\log t$};
    \node(+M-1) [draw, rounded corners, right of=g^M-1, node distance=0.75cm] {$+$};

    \node(L^M-1) [below of=x^M-1, node distance=1cm] {$L^{(M-1)}$};
    \node (x^M) [right of=+M-1, node distance=2cm] {$(\hat{\boldsymbol{s}}_0^{(M)},\log \hat{\boldsymbol{\kappa}}_0^{(M)})$};

    % \node (h^M)  [below of=x^M, node distance=0.75cm] {$\boldsymbol{h}^{(M)}$};
    \node(L^M) [below of=x^M, node distance=1cm] {$L^{(M)}$};

    \draw[arrow, orange] (x^0) to[] (g);
    % \draw[arrow, orange] (h^0) to[bend right] (g);
    \draw[arrow, red] ([xshift=-0.2cm]x^0.north) to[] ([xshift=-0.2cm]y^0.south);
    \draw[arrow, red] ([xshift=-0.1cm,yshift=-0.1cm]y^0.east) to[] ([xshift=0.2cm,yshift=-0.1cm]nabla_x^0.west);
    \draw[arrow, red] ([xshift=-0.1cm,yshift=-0.1cm]y^0.east) to[] ([xshift=0.02cm,yshift=0cm]res^0.west);
    \draw[arrow, red] ([xshift=0.8cm,yshift=0.2cm]nabla_x^0.south) to[bend left] (g);
    \draw[arrow, red] ([xshift=0.1cm,yshift=0.15cm]res^0.south) to[bend left] (g);
    \draw[arrow, orange] ([xshift=-0.22cm,yshift=0.22cm]xt0.east) to (g);
    \draw[arrow, orange] (y0) to (g);
    \draw[arrow, orange] (t0) to (g);
    \draw[arrow, mygreen] (g) to (+0);
    \draw[arrow, orange] ([xshift=1.1cm,yshift=0.2cm]x^0.south) to[bend right] (+0.south);

    \draw[arrow, mygreen] (+0) to[] (x^1);
    % \draw[arrow, mygreen] (g) to[bend right] (h^1);
    
    \draw[arrow, mygreen] (x^1) to[] (g1);
    % \draw[arrow, mygreen] (h^1) to[bend right] (g1);
    \draw[arrow, red] ([xshift=-0.2cm]x^1.north) to[] ([xshift=-0.2cm]y^1.south);
    \draw[arrow, red] ([xshift=-0.1cm,yshift=-0.1cm]y^1.east) to[] ([xshift=0.2cm,yshift=-0.1cm]nabla_x^1.west);
    \draw[arrow, red] ([xshift=-0.1cm,yshift=-0.1cm]y^1.east) to[] ([xshift=0.02cm,yshift=0cm]res^1.west);
    \draw[arrow, red] ([xshift=0.8cm,yshift=0.2cm]nabla_x^1.south) to[bend left] (g1);
    \draw[arrow, red] ([xshift=0.1cm,yshift=0.15cm]res^1.south) to[bend left] (g1);
    \draw[arrow, orange] ([xshift=-0.22cm,yshift=0.22cm]xt1.east) to (g1);
    \draw[arrow, orange] (y1) to (g1);
    \draw[arrow, orange] (t1) to (g1);

    \draw[arrow, mygreen] (g1) to[] (x^M-1);  
    % \draw[arrow, mygreen] (g1) to[bend right] (h^M-1);
    \draw[arrow, mygreen] ([xshift=-0.2cm]x^1.south) to [] ([xshift=-0.2cm]L^1.north);

    \draw[arrow, mygreen] (x^M-1) to[] (g^M-1);
    % \draw[arrow, mygreen] (h^M-1) to[bend right] (g^M-1);
    \draw[arrow, red] ([xshift=-0.4cm]x^M-1.north) to[] ([xshift=-0.4cm]y^M-1.south);
    \draw[arrow, red] ([xshift=-0.1cm,yshift=-0.1cm]y^M-1.east) to[] ([xshift=0.2cm,yshift=-0.1cm]nabla_x^M-1.west);
    \draw[arrow, red] ([xshift=-0.1cm,yshift=-0.1cm]y^M-1.east) to[] ([xshift=0.02cm,yshift=0cm]res^M-1.west);
    \draw[arrow, red] ([xshift=1.2cm,yshift=0.2cm]nabla_x^M-1.south) to[] (g^M-1);
    \draw[arrow, red] ([xshift=-0.1cm,yshift=0.15cm]res^M-1.south) to[bend left] (g^M-1);
    \draw[arrow, orange] ([xshift=-0.22cm,yshift=0.22cm]xtM-1.east) to (g^M-1);
    \draw[arrow, orange] (yM-1) to (g^M-1);
    \draw[arrow, orange] (tM-1) to (g^M-1);
    \draw[arrow, mygreen] (g^M-1) to (+M-1);
    \draw[arrow, mygreen] ([xshift=1.6cm,yshift=0.2cm]x^M-1.south) to[bend right] (+M-1.south);

    \draw[arrow, mygreen] ([xshift=-0.45cm]x^M-1.south) to [] ([xshift=-0.45cm]L^M-1.north);
    \draw[arrow, mygreen] (+M-1) to (x^M);

    % \draw[arrow, orange] (g^M-1) to[bend right] (h^M);
    \draw[arrow, mygreen] ([xshift=-0.25cm]x^M.south) to[] ([xshift=-0.25cm]L^M.north);
    
  \end{tikzpicture}
    \caption{Unrolled computation graph of the DiRIM. During training, loss gradients with respect to the weights are backpropagated through the green edges only. The orange edges do not contribute to the gradient of the loss function with respect to model parameters. Gradients are not backpropagated through the red edges to avoid instabilities from log-domain backpropagation.}
    \label{fig:Unrolled graph}
\end{figure*}
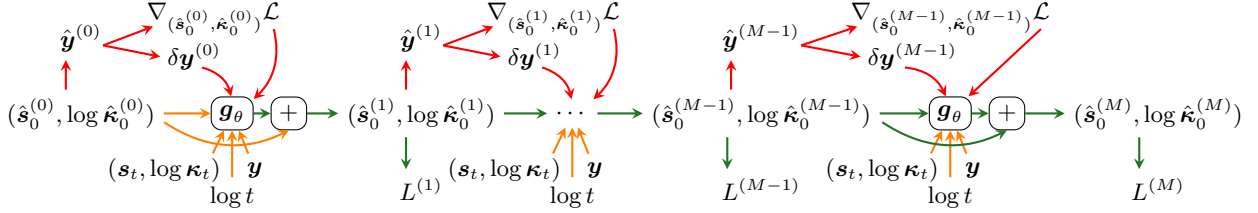

\begin{table}
    \centering
    \caption{RIM hyperparameters used in this work.}
    \begin{tabular}{c||c}
        \hline \hline
        Hyperparameter & Value \\
        \hline
        Number of RIM iterations ($M$) & 5\\
        Iteration loss weights ($w_m$) & $\{0,\frac{1}{4},\frac{1}{4},\frac{1}{4},\frac{1}{4}\}$\\
        Gradients normalization ($c_s$, $c_\kappa$) & $(100,100)$\\
    \end{tabular}
    \label{tab:hyperparameters rim}
\end{table}

During training, the weights of the neural network $\boldsymbol{g}_\theta$ are updated using backpropagation through time (BPTT). Some sections of the computation graph are not recorded for BPTT to avoid instabilities from log-domain backpropagation. This is shown in the full unrolled computation graph in Figure \ref{fig:Unrolled graph}. In addition, we use a reduced number of RIM iterations $(M=2)$ for the first few epochs of training, since failure to do so resulted in instabilities during some of our experiments. 

During training, the source images and convergence maps from the training sets are not paired; instead, they are shuffled at every epoch. Additionally, both source and convergence map images are randomly rotated and mirrored during training. Together, these two choices prevent overfitting. Once the learning rate has decayed to $1\times 10^{-6}$, we fix the learning rate and turn on exponential moving average (EMA) of the weights, and continue training until convergence of the validation loss. Hyperparameters relating to the training of the model are detailed in Table  \ref{tab:hyperparameters training}. 

The training of the model required 72 hours on a single A100 40GB GPU. At inference time, the generation of joint source and kappa map samples required approximately 5 seconds per sample on the same GPU.

\begin{table}
    \centering
    \caption{Training hyperparameters used in this work.}
    \begin{tabular}{c||c}
        \hline \hline
        Hyperparameter & Value \\
        \hline
        Optimizer & Adam\footnote{\cite{Kingma_Ba_2017}}\\
        Initial Learning rate & $2\times 10^{-4}$\\
        Learning rate schedule & Exponential decay\\
        Decay rate per epoch & $0.96$\\
        Batch size & 16\\
        Gradient norm clipping & 10\\
        Weight EMA decay rate per epoch & 0.9999\\
    \end{tabular}
    \label{tab:hyperparameters training}
\end{table}

\section{Additional Reconstructions} \label{app:more results}

\begin{figure}[t]
    \centering
    \includegraphics[width=1\linewidth]{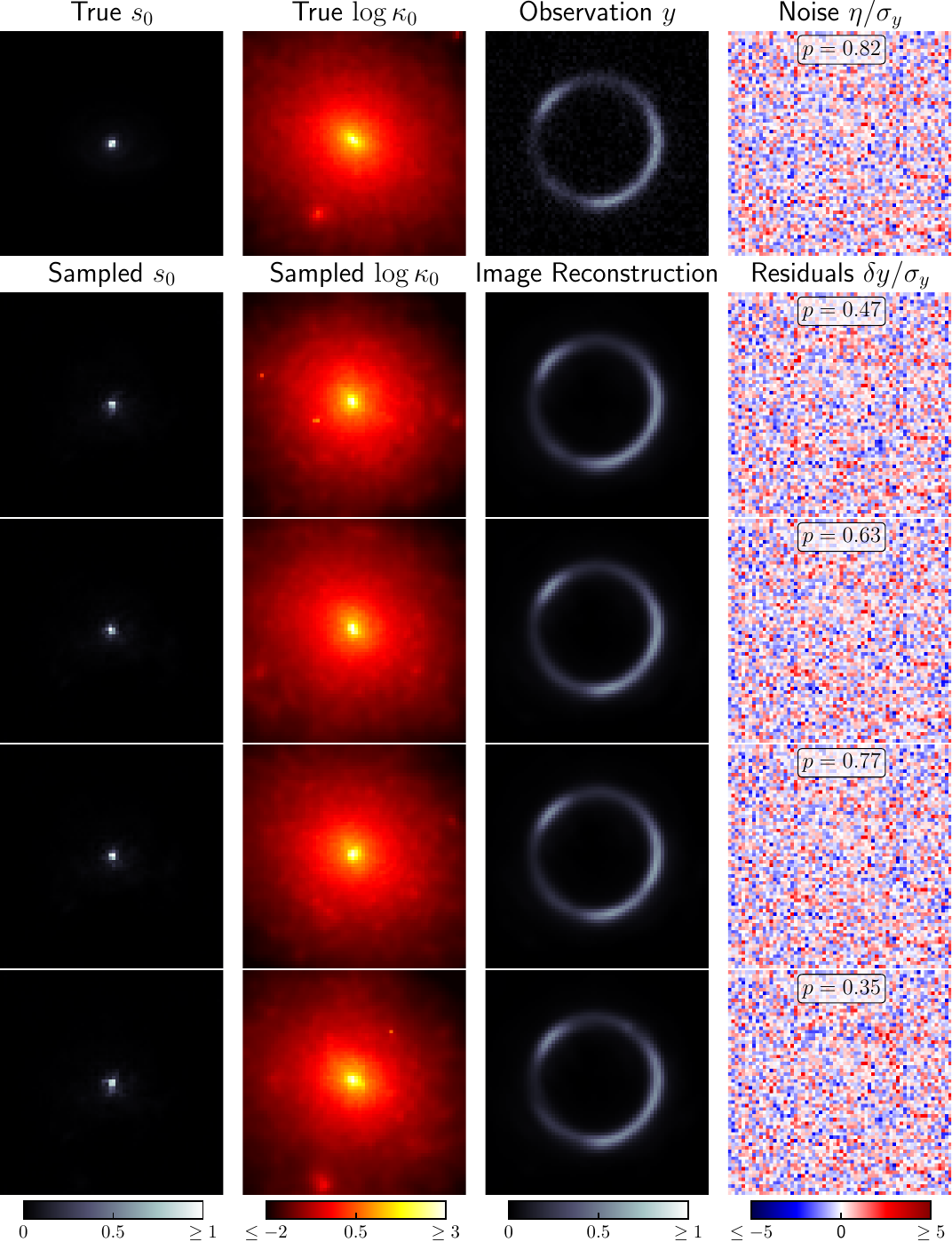}
    \caption{Same as Figure \ref{fig:simulated convergence reconstruction simple} for a randomly selected test set example.}
    \label{fig:simulated convergence reconstruction appendix 1}
\end{figure}

\begin{figure}[t]
    \centering
    \includegraphics[width=1\linewidth]{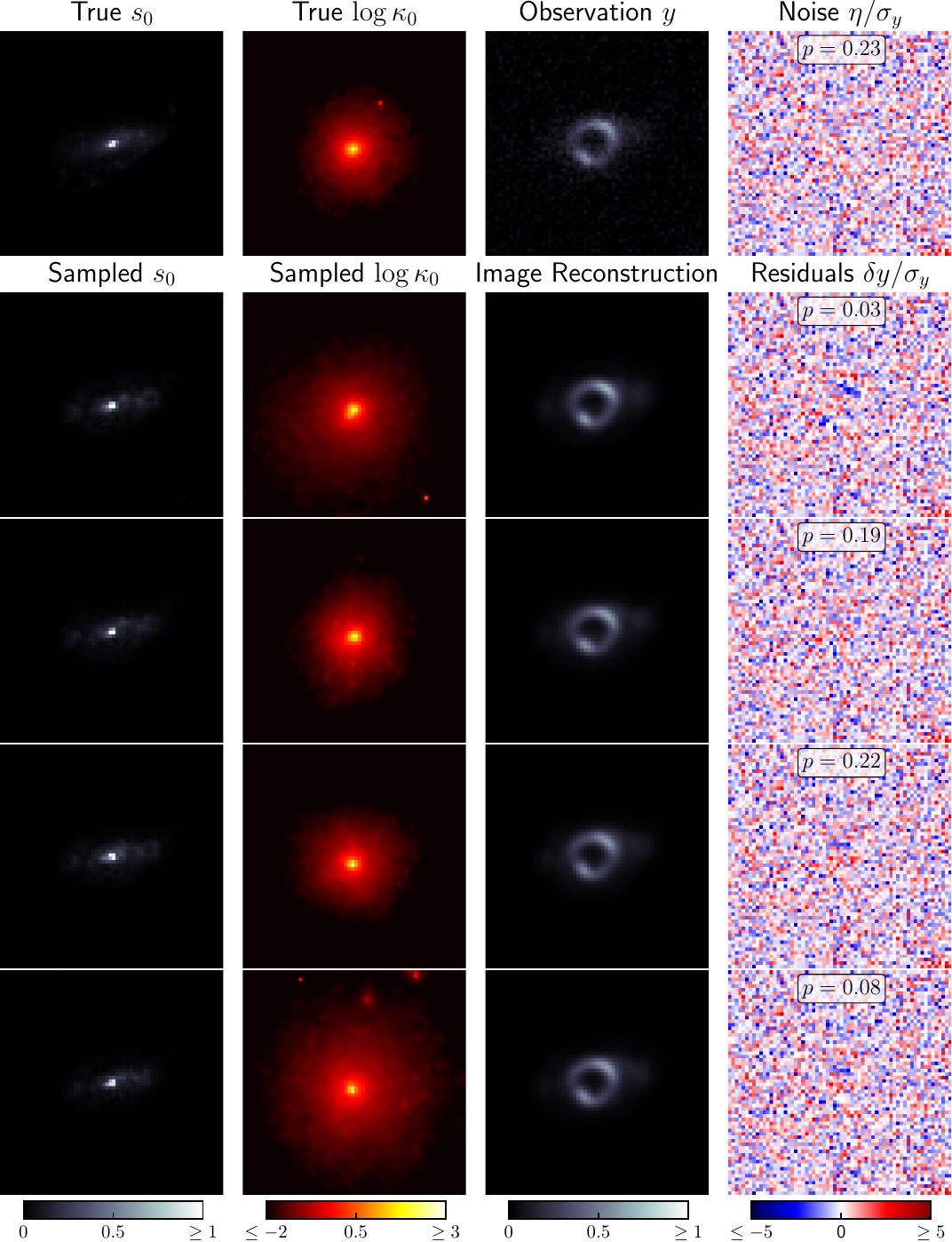}
    \caption{Same as Figure \ref{fig:simulated convergence reconstruction simple} for a randomly selected test set example.}
    \label{fig:simulated convergence reconstruction appendix 10}
\end{figure}

Figures \ref{fig:simulated convergence reconstruction appendix 1}-\ref{fig:simulated convergence reconstruction appendix 10} show ten examples of lens models with observations taken randomly from the test set.

\section{Foreground mass macromodel} \label{app:macromodel}

The macromodel for the foreground mass distribution used in Figure \ref{fig:simulated convergence comparison traditional} to compare our methods to methods modeling the foreground mass distribution using parametric models consists of an EPL main halo with convergence given by Eq.\ref{eq:convergence EPL}, parameterized by the Einstein radius $R_E['']$, the density slope $\tau$  and the axis ratio $q$. In addition, the EPL component is rotated by an angle $\phi$. The EPL main halo is augmented with an external shear, which has zero convergence and is implemented by its reduced deflection angle
\begin{align}
    \boldsymbol{\alpha}(\theta_x,\theta_y) = \gamma \begin{pmatrix}
        \theta_x \cos(2\theta_{\text{ext}}) + \theta_y \sin(2\theta_{\text{ext}})\\
        \theta_x \sin(2\theta_{\text{ext}}) - \theta_y \cos(2\theta_{\text{ext}})
    \end{pmatrix},
\end{align}
parametrized by the external shear amplitude $\gamma$ and its orientation $\theta_{\text{ext}}$. The EPL main halo is also augmented with $m=3$ and $m=4$ multipoles with convergence given by Eq.\ref{eq:convergence multipoles}, where $r =\sqrt{\theta_x^2+\theta_y^2}$ and $\psi = \arctan(\theta_y/\theta_x)$. The multipoles are parameterized by the amplitudes $a_m[('')^{-1}]$ and the orientations $\theta_m$. The EPL, external shear and multipole components are given common center angular coordinates parametrized by $(x_0[''],y_0[''])$. This represents a total of 12 macromodel parameters.

When using this macromodel, the convergence map is not rendered on a grid; instead, the deflection angles are evaluated analytically at the required angular positions during ray tracing. We have, however, also conducted experiments where the convergence map was rendered on a $64\times64$ grid with various levels of upsampling beforehand. This did not improve the fit to the data in Figure \ref{fig:simulated convergence comparison traditional}. This confirms that the poor fit is caused by the lack of flexibility in the foreground mass model.  

\section{RIM Iterations Demonstration} \label{app:demo rim iterations}

\begin{figure*}[!t]
    \centering
    \includegraphics[width=0.9\linewidth]{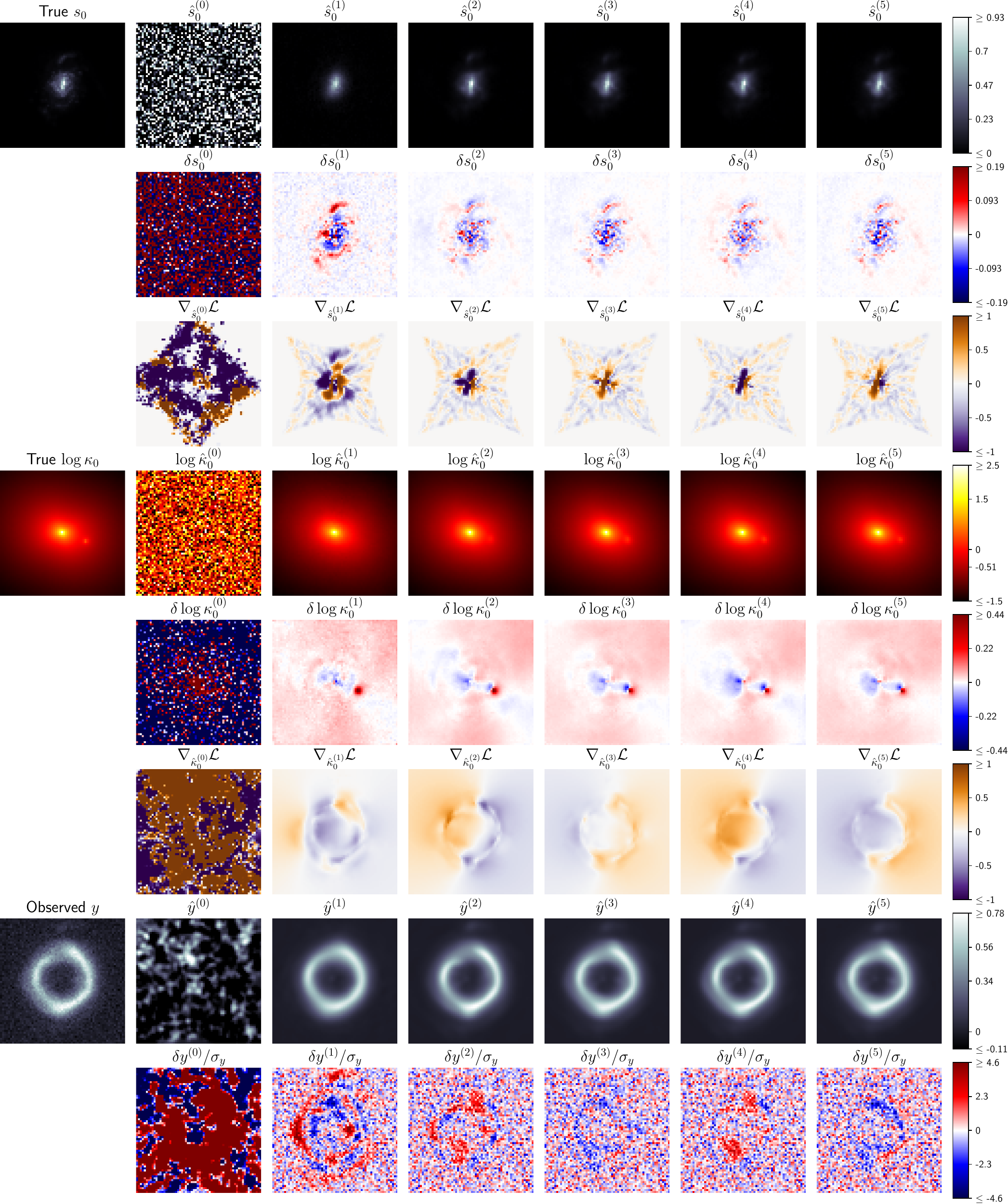}
    \caption{Evolution of the source and convergence map through the $M=5$ denoising RIM iterations at $t=1.0$. This time snapshot is taken from a full solve of the reverse SDE done to generate a joint source and convergence map sample for the same mock observation as in Figure \ref{fig:analytic convergence reconstruction complex}}
    \label{fig:analytic convergence demo rim iterations t=1.0}
\end{figure*}

\begin{figure*}[!t]
    \centering
    \includegraphics[width=0.9\linewidth]{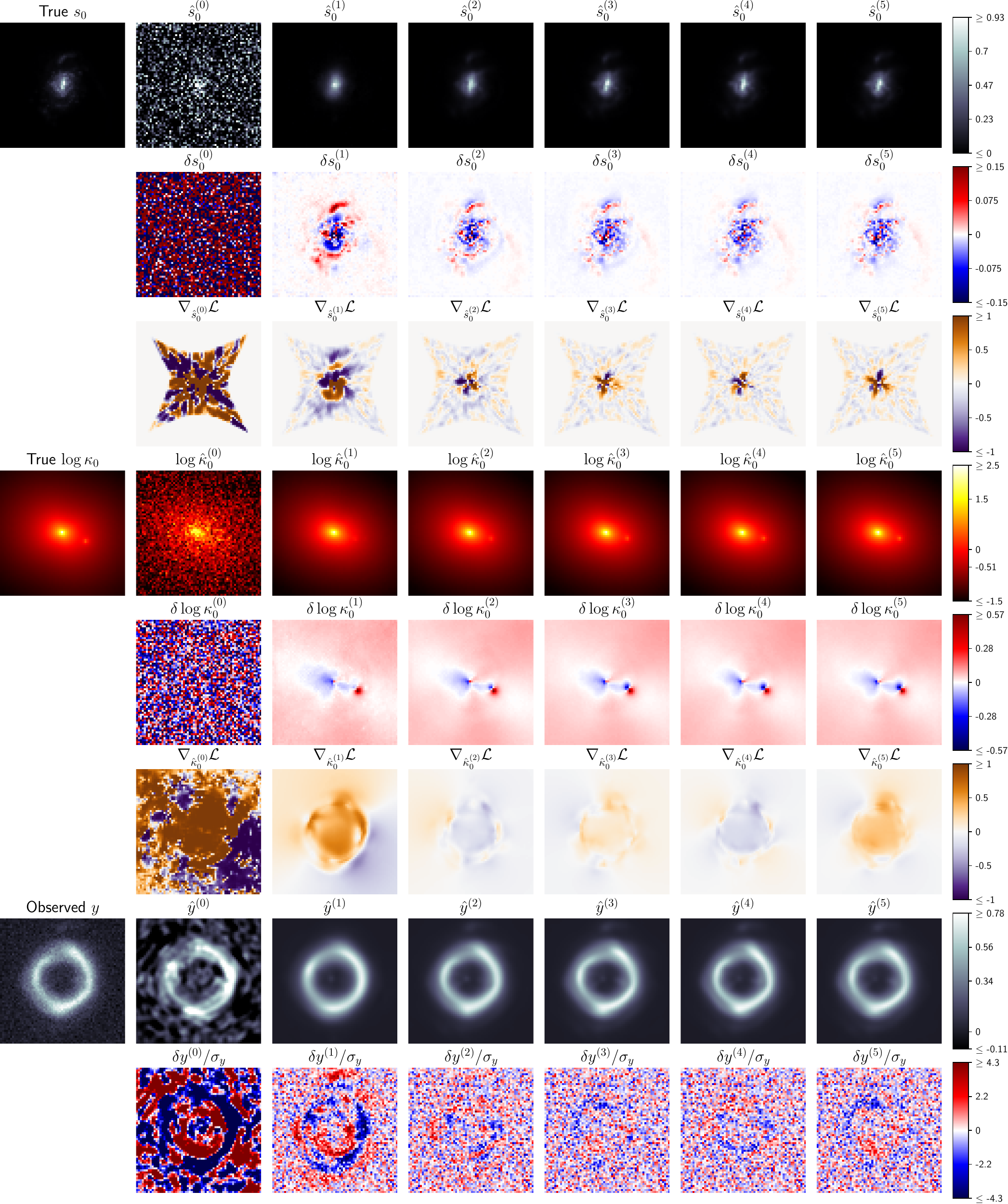}
    \caption{Same as Figure  \ref{fig:analytic convergence demo rim iterations t=1.0} for $t=0.6$.}
    \label{fig:analytic convergence demo rim iterations t=0.6}
\end{figure*}

\begin{figure*}[!t]
    \centering
    \includegraphics[width=0.9\linewidth]{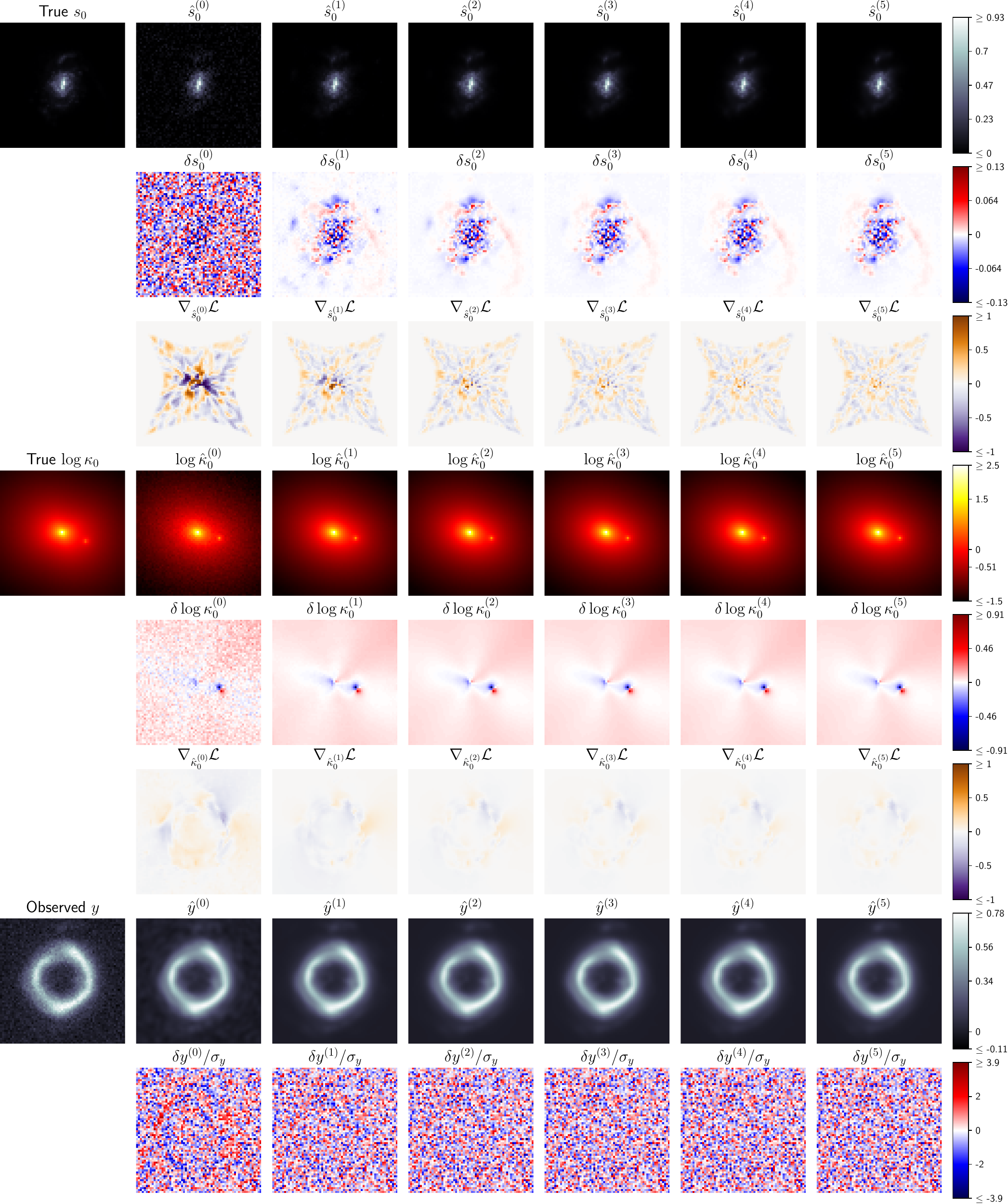}
    \caption{Same as Figure  \ref{fig:analytic convergence demo rim iterations t=1.0} for $t=0.2$.}
    \label{fig:analytic convergence demo rim iterations t=0.2}
\end{figure*}

The posterior scores used to generate the joint source and convergence samples in Figure \ref{fig:analytic convergence reconstruction complex} are generated by a denoising process, as described in Section \ref{sec:dirim}. For visualization purposes, we show in Figure \ref{fig:analytic convergence demo rim iterations t=1.0}, Figure \ref{fig:analytic convergence demo rim iterations t=0.6} and Figure \ref{fig:analytic convergence demo rim iterations t=0.2} the evolution of the source and convergence map through the $M=5$ denoising RIM iterations at $t=1.0$, $t=0.6$ and $t=0.2$ respectively. These time snapshots are taken from a full solving of the reverse SDE done to generate a joint source and convergence map sample as in  Figure \ref{fig:analytic convergence reconstruction complex}. In all three figures, it can be seen that the model predictions for the denoised source and convergence map get progressively better through the RIM iterations. Moreover, at $t=1.0$, it can be seen that the subhalo present in the true convergence map is undetected by the model after the first RIM iteration, and is only found after subsequent RIM iterations. This represents an additional demonstration of the importance of the RIM in our methods.

\newpage

\begin{figure}[t]
    \centering
    \includegraphics[width=1\linewidth]{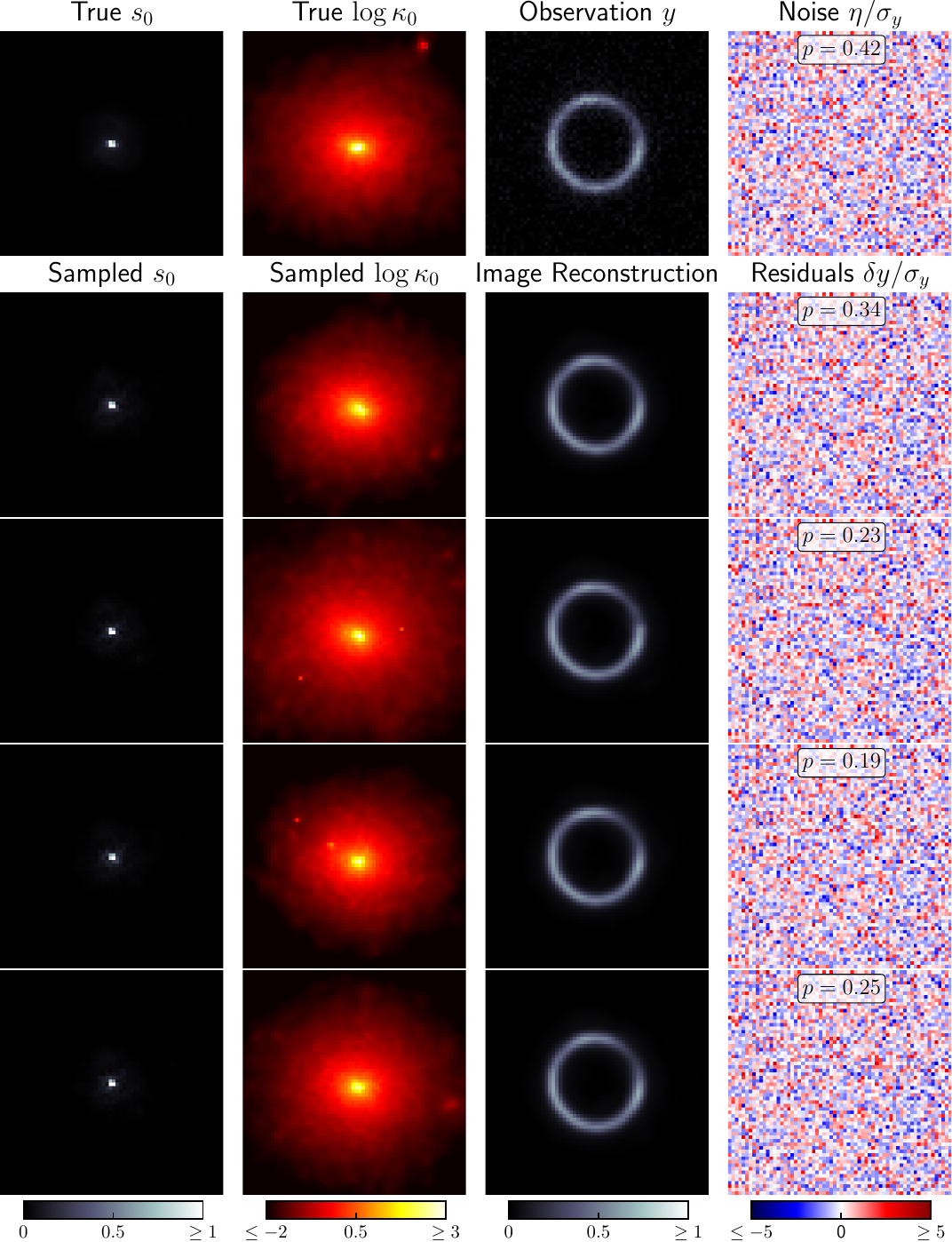}
    \caption{Same as Figure \ref{fig:simulated convergence reconstruction simple} for a randomly selected test set example.}
\end{figure}

\begin{figure}[t]
    \centering
    \includegraphics[width=1\linewidth]{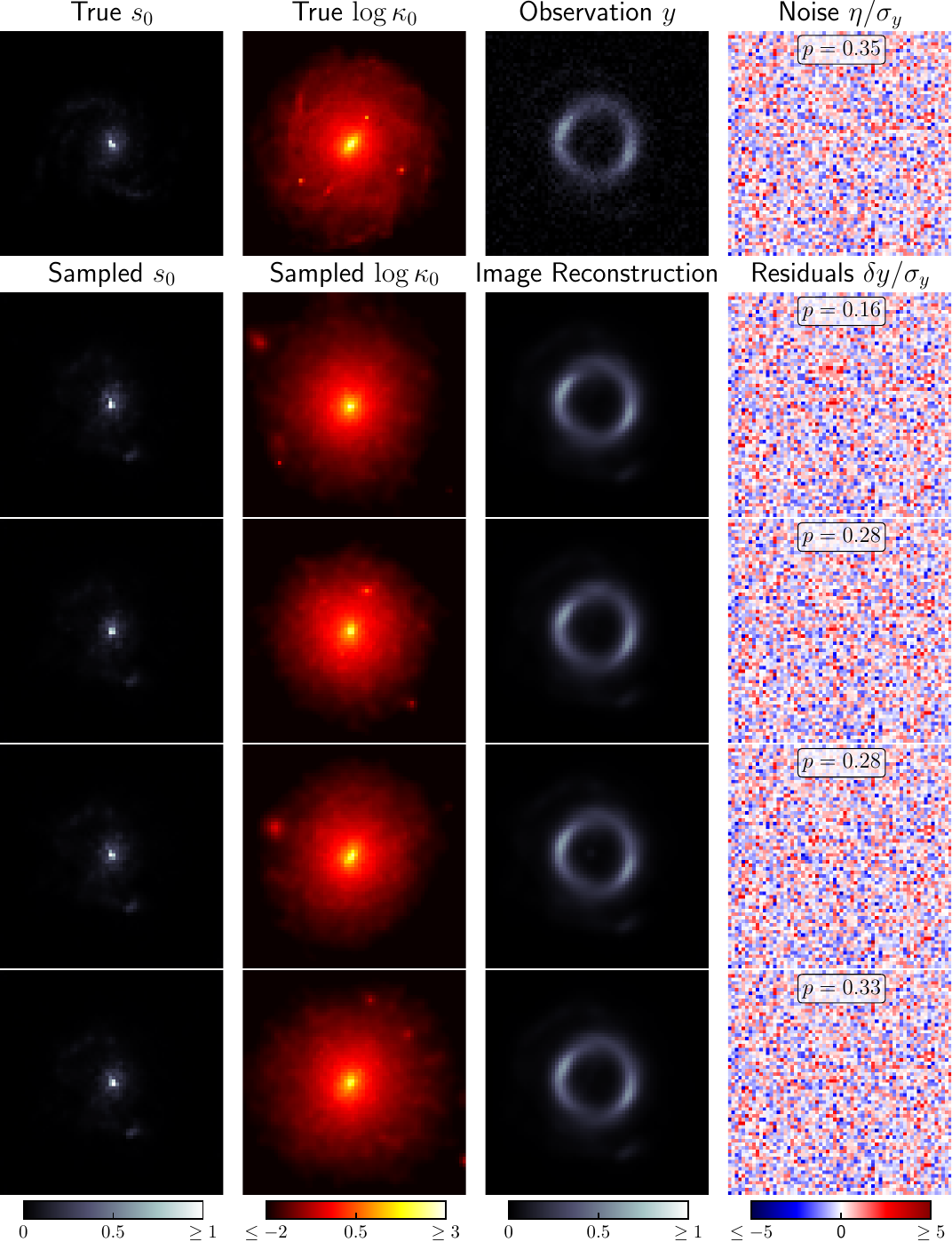}
    \caption{Same as Figure \ref{fig:simulated convergence reconstruction simple} for a randomly selected test set example.}
\end{figure}

\begin{figure}[t]
    \centering
    \includegraphics[width=1\linewidth]{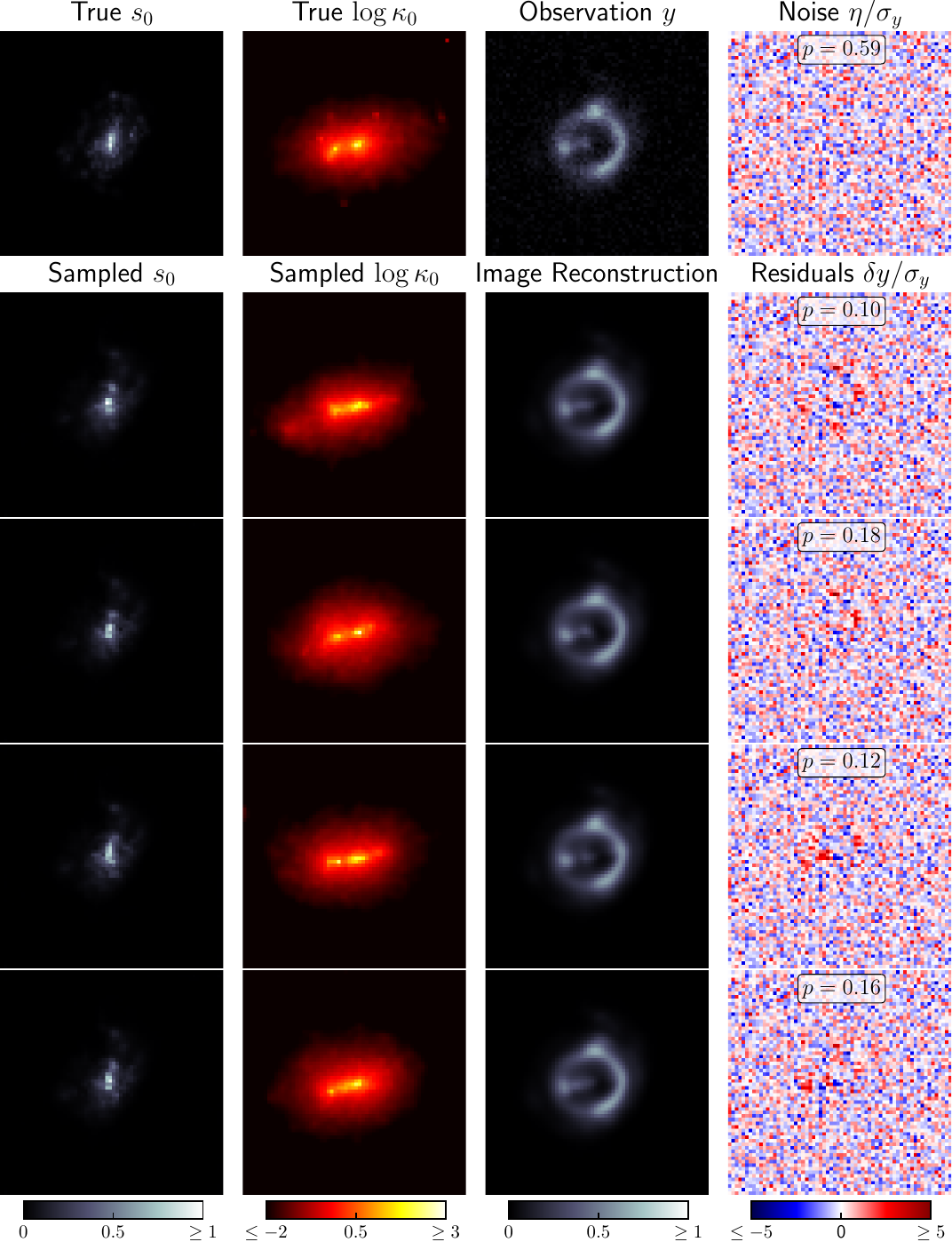}
    \caption{Same as Figure \ref{fig:simulated convergence reconstruction simple} for a randomly selected test set example.}
\end{figure}

\begin{figure}[t]
    \centering
    \includegraphics[width=1\linewidth]{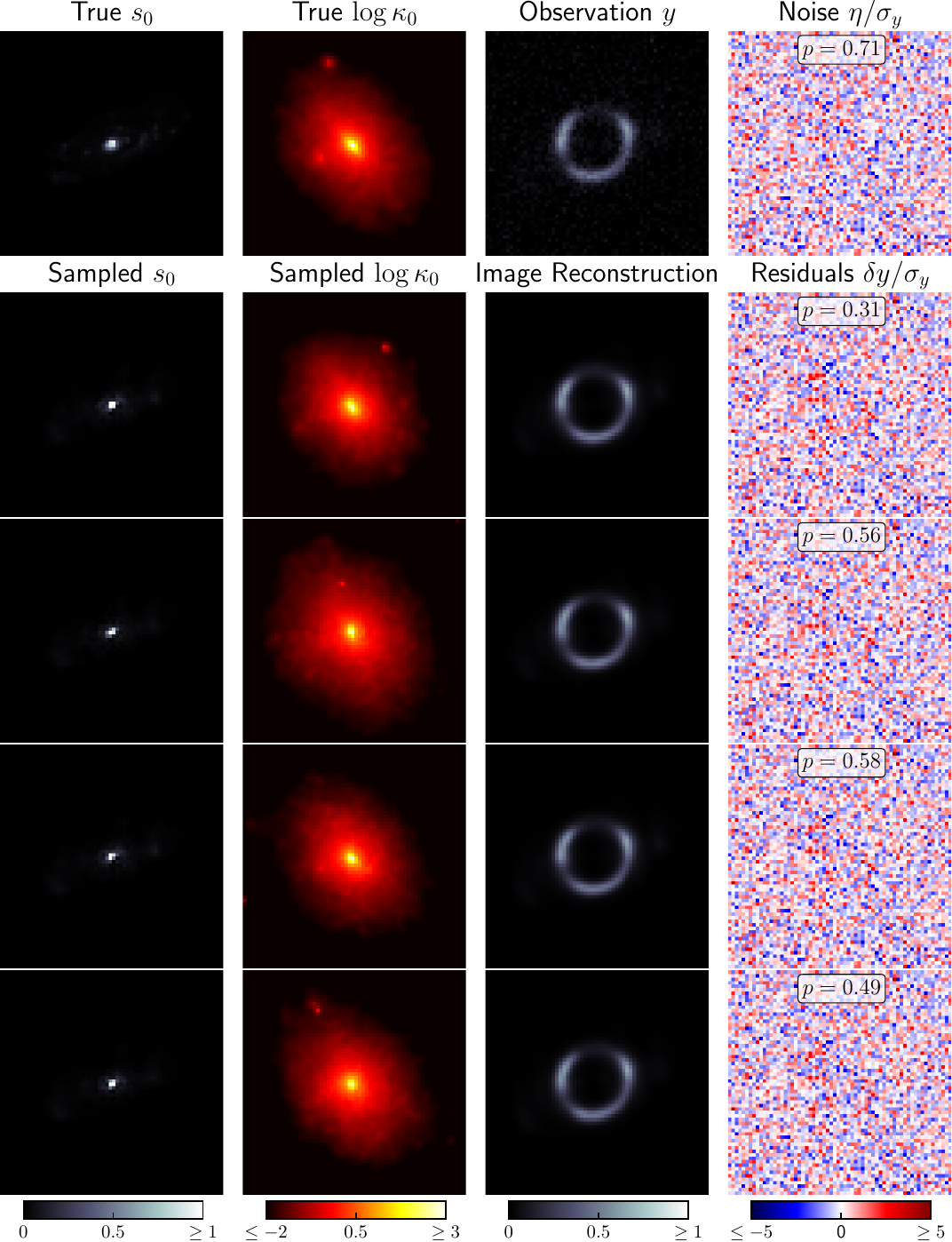}
    \caption{Same as Figure \ref{fig:simulated convergence reconstruction simple} for a randomly selected test set example.}
\end{figure}

\begin{figure}[t]
    \centering
    \includegraphics[width=1\linewidth]{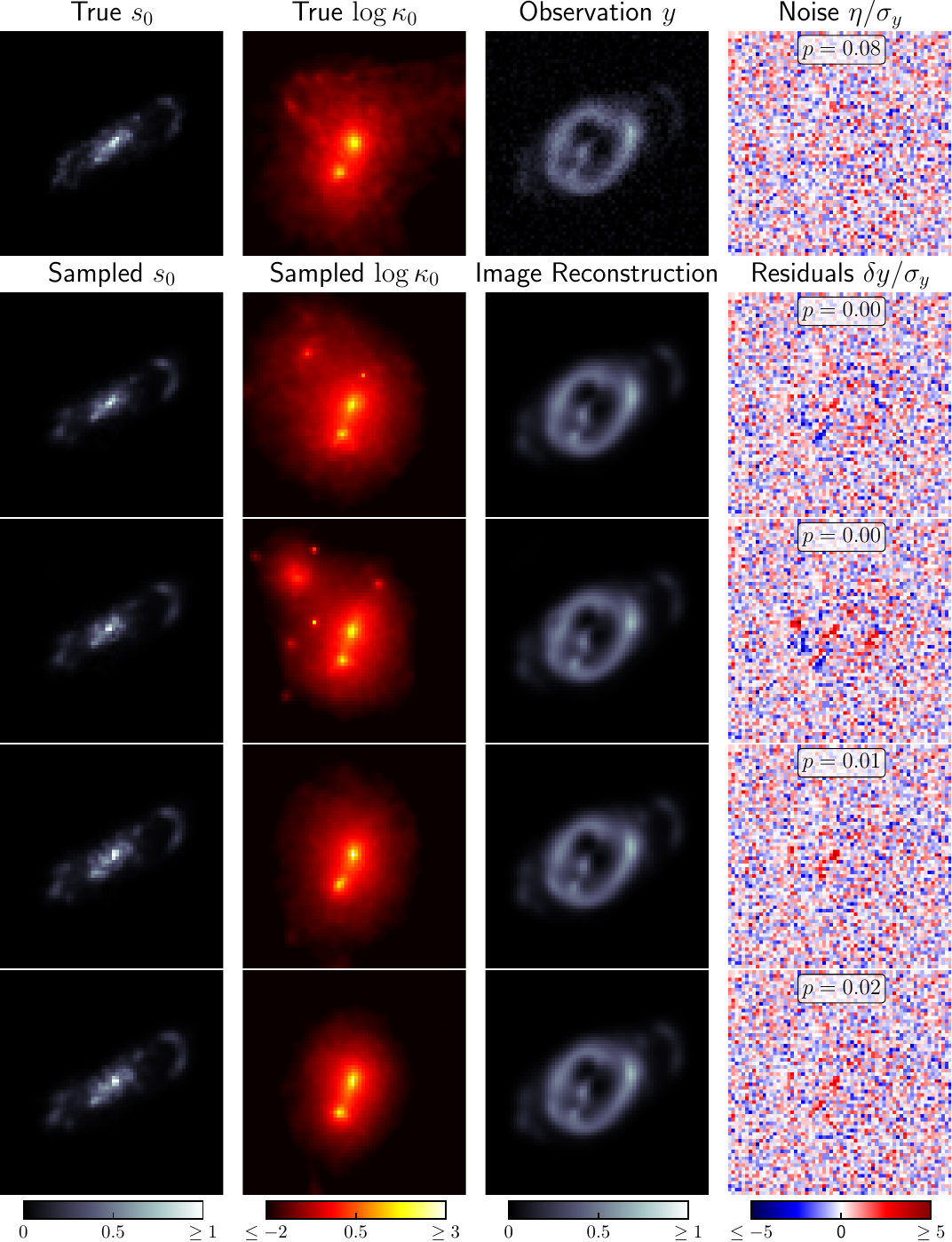}
    \caption{Same as Figure \ref{fig:simulated convergence reconstruction simple} for a randomly selected test set example.}
\end{figure}

\begin{figure}[t]
    \centering
    \includegraphics[width=1\linewidth]{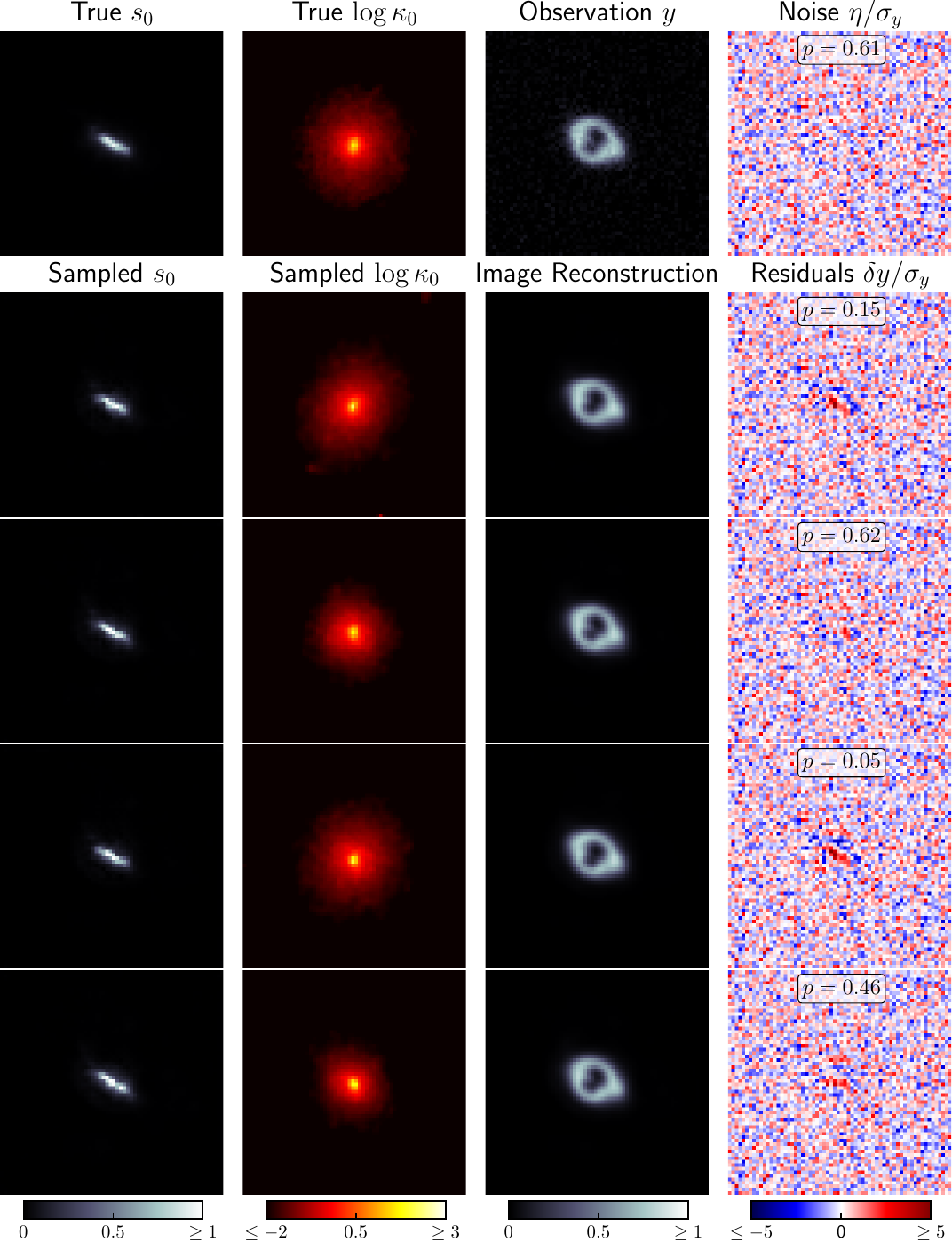}
    \caption{Same as Figure \ref{fig:simulated convergence reconstruction simple} for a randomly selected test set example.}
\end{figure}

\begin{figure}[t]
    \centering
    \includegraphics[width=1\linewidth]{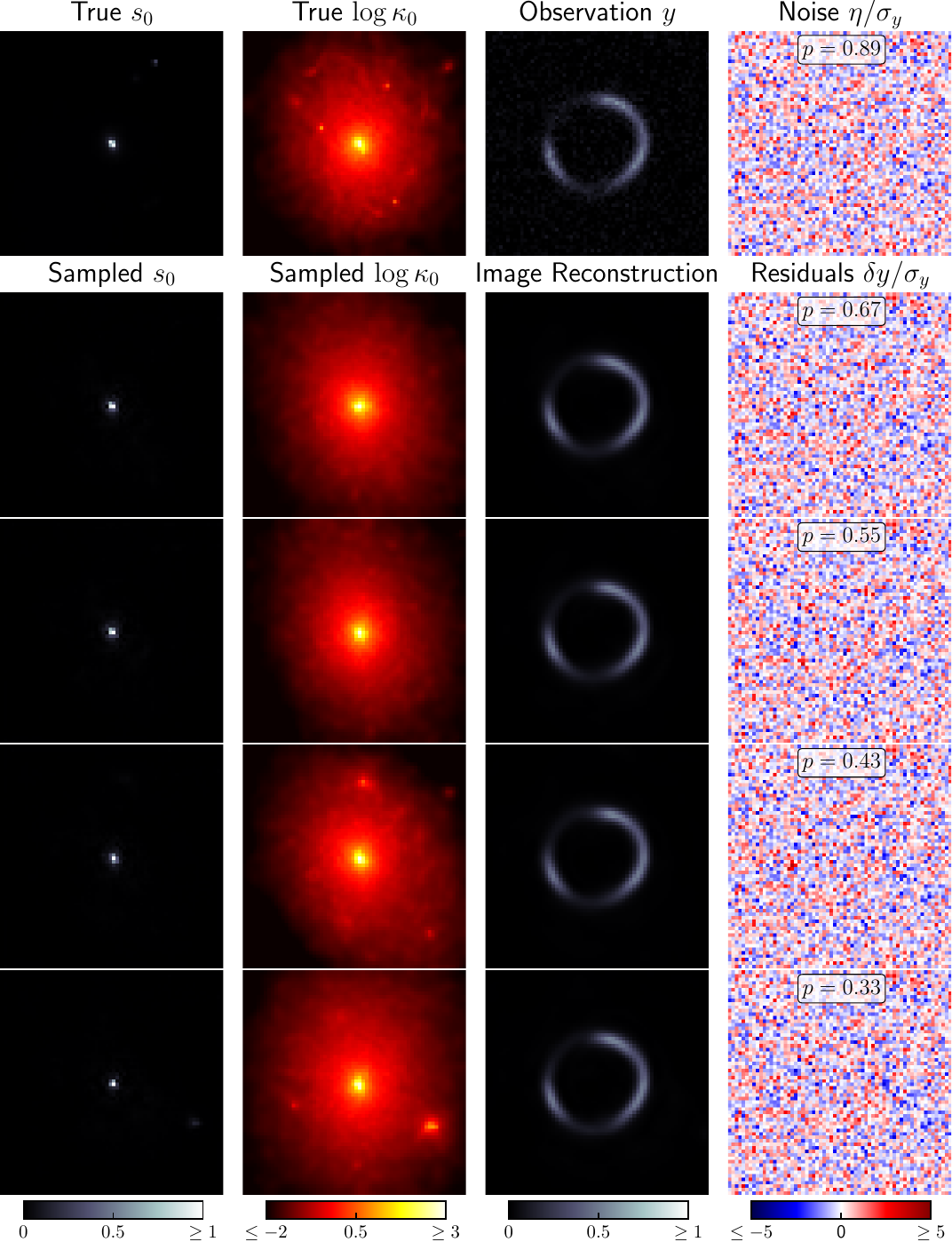}
    \caption{Same as Figure \ref{fig:simulated convergence reconstruction simple} for a randomly selected test set example.}
\end{figure}

\begin{figure}[t]
    \centering
    \includegraphics[width=1\linewidth]{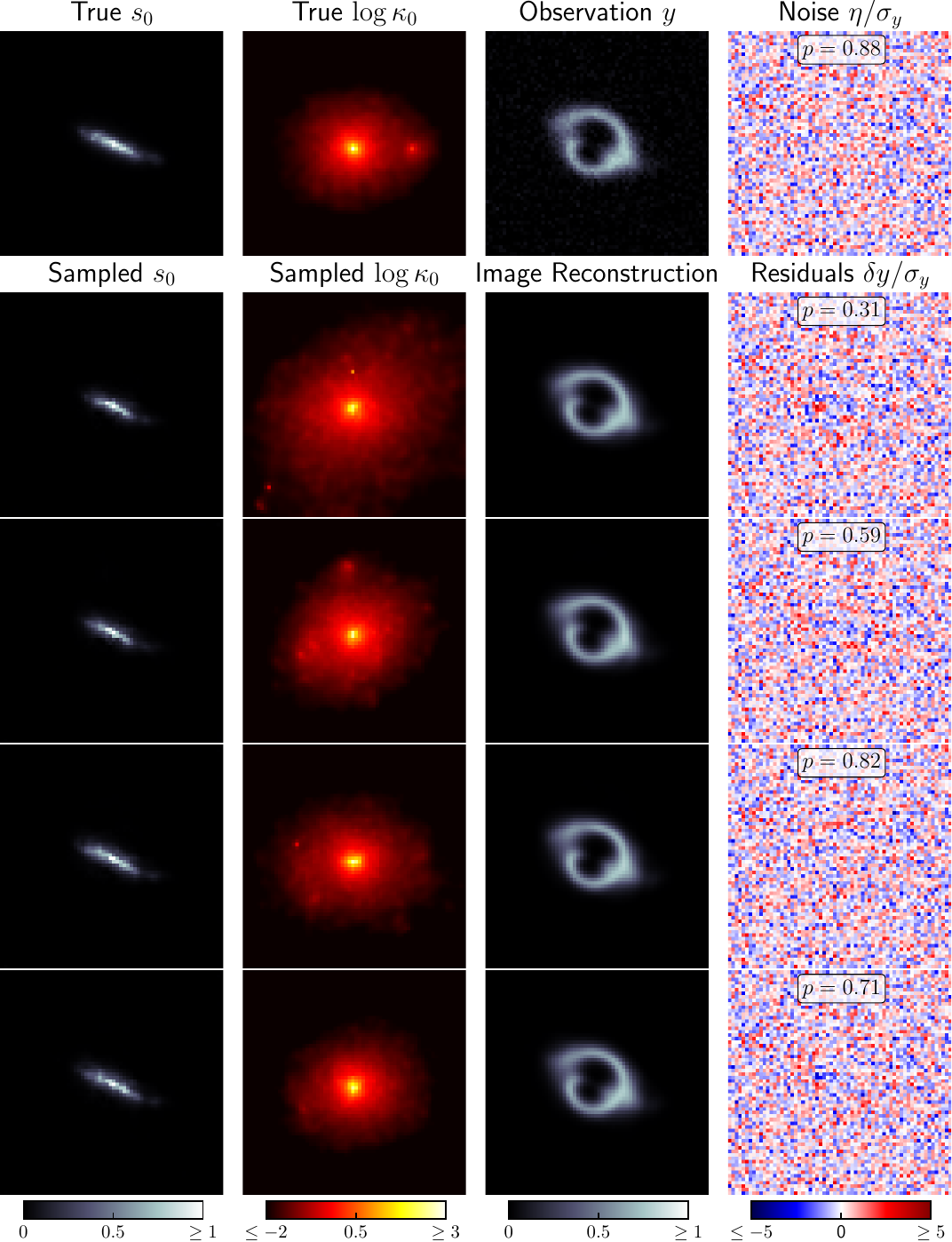}
    \caption{Same as Figure \ref{fig:simulated convergence reconstruction simple} for a randomly selected test set example.}
\end{figure}

\end{document}